\documentclass[%
 aip,
 amsmath,amssymb,
 reprint,%
author-year,%
]{revtex4-1}

\usepackage{graphicx}
\usepackage{dcolumn}
\usepackage{bm}
\usepackage[utf8]{inputenc}
\usepackage[T1]{fontenc}
\usepackage{mathptmx}
\usepackage{gensymb}

\usepackage{dcolumn}
\usepackage[normalem]{ulem}

\newcommand{\Ree}{{\rm Re}}
\newcommand{\Ha}{{\rm Ha}}
\newcommand{\Rm}{{\rm Rm}}
\newcommand{\Pm}{{\rm Pm}}

\newcommand{\vo}{{\mathbf{w}}}

\newcommand{\ez}{{\bf\hat e_z}}
\newcommand{\ve}{{\mathbf{v}}}

\newcommand{\rv}{\ensuremath{\mathbf{r}}}

\newcommand{\lp}{\ensuremath{\left(}}
\newcommand{\rp}{\ensuremath{\right)}}

\usepackage{color}
\definecolor{blue-violet}{rgb}{0.54, 0.17, 0.89}

\begin{document}

\preprint{AIP/123-QED}

\title{Modulated rotating waves and triadic resonances in spherical
  fluid systems: The case of magnetized spherical Couette flow}

\affiliation{
Institute of Fluid Dynamics, Helmholtz-Zentrum Dresden-Rossendorf, Bautzner Landstraße 400, 01328
Dresden, Germany
}

\author{F. Garcia}
\altaffiliation{Author to whom correspondence should be addressed: f.garcia-gonzalez@hzdr.de}
\author{A. Giesecke}
\author{F. Stefani}

\date{\today}

\begin{abstract}
The existence of triadic resonances in the magnetized spherical
Couette system (MSC) is related to the development of modulated
rotating waves, which are quasiperiodic flows understood in terms of
bifurcation theory in systems with symmetry. In contrast to previous
studies in spherical geometry the resonant modes are not inertial
waves but related with the radial jet instability which is strongly
equatorially antisymmetric. We propose a general framework in which
triadic resonances are generated through successive Hopf bifurcations
from the base state.  The study relies on an accurate frequency
analysis of different modes of the flow, for solutions belonging to
two different bifurcation scenarios. The azimuthal and latitudinal
nonlinear coupling among the resonant modes is analysed and
interpreted using spherical harmonics and the results are compared
with previous studies in spherical geometry.
\end{abstract}

\maketitle


\section{Introduction}
\label{sec:Int}

The onset of instabilites in rotating fluids is often related to the
emergence of a triadic resonance, which describes a resonant
interaction between three internal modes or waves that requires the
fulfillment of certain selection rules in order to take place.  Such
interactions occur on a small scales in the laboratory as well as on
large scales in geo- or astrophysically relevant fluid flows.  In the
case of ocean flows, for example, the relevance of weak wave resonant
interactions for the global wave evolution has been investigated for
several decades (e.\,g. \cite{HaHe93}). More recent examples comprise
the study of internal tides models of \cite{SuJe20}, the experiments
on gravity surface turbulence of \cite{CHRVVSM19}, or the theoretical
study of \cite{LLLB20} devoted to the analysis of steady state
resonant interfacial waves in a two layer fluid filling a duct.

Regarding celestial bodies, rotating (and electrically conducting)
fluids are responsible for the generation of magnetic fields
(\cite{Jon11}).  These flows are assumed to be turbulent which raises
the question of the initial instabilities and the related weakly
nonlinear interactions (e.\,g. \cite{BCL15}).  It is well known that
resonant interactions are responsible for energy transfers among
different modes (e.\,g. \cite{Wal92,SmWa99,Gal03}). Recent experiments
of \cite{BGC20} on wave-driven rotating turbulence have shown the
existence of a secondary instability involving four resonant waves
that feed an unforced geostrophic flow.

The triadic instabilites  found in simulations and
experiments of precessing cylinders (\cite{LMNE11,HGGS15,GAGHS15}) 
have been linked to Hopf bifurcations from the steady base state that
goes along with the transition from regular (periodic and
quasiperiodic) flows to chaotic flows (\cite{ABLMM15,LoMa18}).

Recently, triadic resonances were also identified in spherical systems
(e.\,g. \citet{HHT16,BTHW18,Lin21}). In the case of mechanically
driven Couette flow in a wide-gap spherical shell, \citet{BTHW18}
discussed the appearance of a triadic resonance regime in which free
inertial modes form a triad with the basic non-axisymmetric
instability. For convection driven fluid flows in full spheres a
triadic resonance regime involving inertial waves was found in
\cite{Lin21} linking convection and mechanical drivings in agreement
with the theoretical analysis of \cite{ZhLi17} performed in terms of
inertial waves and for either convective or mechanically forced
rotating fluids systems.

Inertial waves are solutions of the inviscid Navier-Stokes equations
for which the Coriolis effects are dominant
(e.\,g. \cite{Gre68,RiVa97,ZELB01}). In the context of the spherical
Couette flow inertial wave-like motions were identified in the
experimental study of \cite{KTZTL07}, which provided evidence of the
generation of these type of motions when the flow is driven by strong
differential rotation.  For the same problem, the study of
\citet{BTHW18} showed that at higher rotation rates the fundamental
instability is a shear-layer instability (\cite{Ste66,HFME04}),
equatorially symmetric and nearly geostrophic, which forms a triad
with two equatorially antisymmetric inertial modes. This instability
emerges when Coriolis forces prevail (as required for inertial
geostrophic columns, see \cite{Wic14}).

We demonstrate that the weakly nonlinear solutions of the magnetized
spherical Couette system (MSC) satisfy triadic resonance interactions
that involve different azimuthal and latitudinal modes. These modes
are not inertial waves but related to the radial jet instability
(\cite{HJE06,GaSt18,GSGS19,GSGS20b}), which is basically hydrodynamic
because the effects of the applied magnetic field are
weak. Contrasting to the results of \citet{BTHW18} the radial jet
instability develops when viscous, rather than Coriolis, forces are
dominant. Triadic resonant flows emerge from the base state via a
sequence of Hopf bifurcations similar to the scenario described in
\cite{ABLMM15} and \cite{LoMa18} for the precessing cylinder. We show
that this link between triadic instability and Hopf bifurcations to
periodic and quasiperiodic flows is then valid as well in spherical
geometry using the particular example of the MSC system. The sequence
of bifurcations is obtained by modifying the magnetic field strength,
though similar bifurcations are expected when the magnitude of imposed
differential rotation is increased (\cite{Hol09,TEO11}).

Concretely, the quasiperiodic solutions having resonant modes are
modulated rotating waves (MRW), for which a developed theory in terms
of equivariant dynamical systems exists
(\cite{Ran82,CrKn91,CoMa92,GLM00,GoSt03}). These solutions have
certain spatio-temporal symmetries and appear since the system is
{\bf{SO}}$(2)\times${\bf{Z}}$_2$-equivariant, i.\,e., invariant by
azimuthal rotations and reflections with respect to the equatorial
plane. We show that an instability in terms of a triadic resonance
consisting of three resonantly interacting modes can equivalently be
described in terms of a MRW.  The spatio-temporal symmetry of the MRWs
renders possible the existence of triadic resonant relations among
several modes which reflect the quadratic nature of the nonlinear
interaction.

While the mechanism proposed here is more general, it is not in
contradiction with the interpretation of triadic resonances in terms
of inertial waves, as quasiperiodic MRW involving inertial modes can
be found as well in spherical (e.\,g. \cite{GCW19,Lin21}) and
cylindrical geometry (e.\,g. \cite{ABLMM18}), and their mathematical
description is analogous to that of inertial waves of \cite{BTHW18}
and the references therein. We demonstrate that resonant interactions
prevail on a significant region of the parameter space and even for
chaotic flows, generated from further bifurcations of quasiperiodic
MRW.

The organization of the paper is the following: In \S\ \ref{sec:mod}
the model problem is introduced and the numerical method used is
briefly described. In \S\ \ref{sec:triad_reson} we argue how triadic
resonances develop for MRWs considering the azimuthal dependence. The
identification and description of resonant triads for two different
scenarios and the comparison with previous studies is completed in
\S\ \ref{sec:res} while the latitudinal dependence among the resonant
modes is discussed in \S\ \ref{sec:nlin_colat}.  The paper closes with
\S\ \ref{sec:conc}, summarizing the results obtained.

\section{The model equations}
\label{sec:mod}

The magnetized spherical Couette system consist of an electrically
conducting fluid of density $\rho$, kinematic viscosity $\nu$,
magnetic diffusivity $\eta=1/(\sigma\mu_0)$ (where $\mu_0$ is the
magnetic permeability of the free-space and $\sigma$ is the electrical
conductivity), which fills the space between two spheres with radius
$r_{\text{i}}$ and $r_{\text{o}}$, respectively. The gap width
selected for this study is $\chi=r_{\text{i}}/r_{\text{o}}=0.5$. The
outer sphere is at rest and the inner sphere is rotating at angular
velocity $\Omega$ around the vertical axis $\ez$, and a uniform axial
magnetic field of amplitude $B_0$ is applied. This model has been
successfully employed to compare with the very recent experiments of
\cite{OGGSS20} in which a liquid metal within a differentially
rotating spherical vessel is subjected to an axial magnetic field. In
these experiments RWs of azimuthal symmetries $m\in\{2,3,4\}$ were
found in the return flow instability regime for $\Ree=10^3$ and
$\Ha\in[27.5,40]$.

The system is described by the Navier-Stokes and induction equations
scaling the length, time, velocity and magnetic field with
$d=r_{\text{o}}-r_{\text{i}}$, $d^2/\nu$, $r_{\text{i}}\Omega$ and
$B_0$, respectively. The time dependent equations are:
\begin{align}
& \partial_t\ve+\Ree\lp\ve\cdot\nabla\rp\ve = -\nabla
  p+\nabla^2\ve+\Ha^2(\nabla\times {\bf b})\times\ez , \label{eq:mom}   \\ &
  0 = \nabla\times(\ve\times\ez)+\nabla^2{\bf b}, \quad
  \nabla\cdot\ve=0, \quad \nabla\cdot{\bf b}=0, \label{eq:div}  
\end{align}
where $\Ree=\Omega r_{\text{i}} d/\nu$ is the Reynolds number,
$\Ha=B_0d(\sigma/(\rho\nu))^{1/2}$ is the Hartmann number, $\ve$ the
velocity field and ${\bf b}$ the deviation of the magnetic field from
the axial applied field.

The inductionless approximation is used since it is valid in the limit
of small magnetic Reynolds number, $\Rm=\Omega r_{\text{i}} d/\eta \ll
1$. This is the case for the liquid metal considered, GaInSn, with
magnetic Prandtl number $\Pm=\nu/\eta\sim O(10^{-6})$
(\cite{PSEGN14}), at moderate $\Ree=10^3$, since
$\Rm=\Pm\hspace{0.5mm}\Ree \sim 10^{-3}$. The boundary conditions for
the velocity field are no-slip ($v_r=v_\theta=v_\varphi=0$) at
$r=r_{\text{o}}$ and constant rotation ($v_r=v_\theta=0,~v_\varphi=
\sin{\theta}$, $r$, $\theta$ and $\varphi$ being the radius,
colatitude and longitude, respectively) at $r=r_{\text{i}}$. For the
magnetic field insulating boundary conditions are applied in
accordance with the experimental setups (\cite{SMTHDHAL04,OGGSS20}).

To solve the model equations the toroidal-poloidal formulation
(\cite{Cha81}), in which the velocity field is expressed in terms of
toroidal, $\Psi$, and poloidal, $\Phi$, potentials
\begin{equation}
  \ve=\nabla\times\lp\Psi\rv\rp+\nabla\times\nabla\times\lp\Phi\rv\rp,
\end{equation}
is used. The potentials are expressed in terms of the spherical
harmonics angular functions (truncated at order and degree
$m=l=L_{\text{max}}$) and in the radial direction a collocation method
on a Gauss--Lobatto mesh of $N_r$ points is employed. The toroidal
($\Psi$) and poloidal ($\Phi$) potentials are then
\begin{eqnarray}
  \Psi(t,r,\theta,\varphi)=\sum_{l=0}^{L_{\text{max}}}\sum_{m=-l}^{l}{\Psi_{l}^{m}(r,t)Y_l^{m}(\theta,\varphi)},\label{eq:serie_psi}\\
  \Phi(t,r,\theta,\varphi)=\sum_{l=0}^{L_{\text{max}}}\sum_{m=-l}^{l}{\Phi_{l}^{m}(r,t)Y_l^{m}(\theta,\varphi)},\label{eq:serie_phi}
\end{eqnarray}
with $\Psi_l^{-m}=\overline{\Psi_l^{m}}$,
$\Phi_l^{-m}=\overline{\Phi_l^{m}}$, $\Psi_0^0=\Phi_0^0=0$ to uniquely
determine the two potentials, and
$Y_l^{m}(\theta,\varphi)=P_l^m(\cos\theta) e^{im\varphi}$, $P_l^m$
being the normalized associated Legendre functions of degree $l$ and
order $m$. The code is parallelized in the spectral and in the
physical space by using OpenMP directives. We use optimized libraries
(FFTW3 \cite{FrJo05}) for the FFTs in $\varphi$ and matrix-matrix
products (dgemm GOTO \cite{GoGe08}) for the Legendre transforms in
$\theta$ when computing the nonlinear terms (see \cite{GNGS10} for
further details).

The time integration is based on high order implicit-explicit backward
differentiation formulas (IMEX--BDF) fully described in
\cite{GNGS10}. The nonlinear terms are treated explicitly to avoid the
implicit solution of nonlinear systems. To simplify the resolution of
linear systems, the Lorenz term is also treated explicitly. This may
lead to reduced time steps in comparison with an implicit treatment if
large $\Ha$ numbers are considered, which however is not the case of
the present study with $\Ha\sim O(1)$.

\section{Triadic resonances}
\label{sec:triad_reson}

The conditions for triadic resonances among inertial modes were
established in \citet{BTHW18} (see their sections 2 and 7, and
appendix A). Expanding the velocity and pressure fields in terms of
the Rossby number the zeroth order of the Navier-Stokes equations
leads to the inertial mode equation, with the resonant modes being of
the form
$u_i(t,r,\theta,\varphi)=c_i(t,r,\theta)\text{e}^{i(m_i\varphi-\omega_it)}$,
$i=1,2$. The subsequent analysis of the first order equation provides
the resonance conditions, which are obtained by computing the
quadratic advection term of the sum $u_1+u_2$ (assumed to be an
approximation of the zeroth-order solution).

The time dependence of a rotating wave (RW) is described by a solid
body rotation of a fixed spatial pattern with a certain rotation
frequency $\omega$. Because of this, they are steady flows if the
change of variables $\overline{\varphi}=\varphi+\omega t$ is applied
to the Navier-Stokes equations (Eqs. (\ref{eq:mom})-(\ref{eq:div})),
see \cite{Ran82,SGN13,GNS16,GSGS19} for a theoretical description and
applications of RWs in spherical systems. If the azimuthal symmetry of
the RW is $m_0$
\footnote{If the flow has $m_0$-fold azimuthal symmetry the toroidal
  and poloidal amplitudes of
  Eqs.(\ref{eq:serie_psi},\ref{eq:serie_phi}) are nonzero only for
  $m$'s which are multiples of $m_0$.}
, the spherical harmonic amplitudes of the toroidal and poloidal
potentials (Eqs. (\ref{eq:serie_psi})-(\ref{eq:serie_phi})) become
\begin{eqnarray*}
\Psi_{l}^{m_0k}(r,t)=\overline{\Psi}_{l}^{m_0k}(r)~e^{-im_0k\omega t},\\
\Phi_{l}^{m_0k}(r,t)=\overline{\Phi}_{l}^{m_0k}(r)~e^{-im_0k\omega t},
\end{eqnarray*}
with $k\in\mathbb{Z}$. This is the same expression as given in
Eq. (A5a,b) of \citet{BTHW18} for the inertial modes, restricted to
$m=km_0$. The relation between their values of the frequencies
($\omega_j$) of each inertial mode and the rotation frequency of the
RW waves is $\omega_j=j\omega$. The above expressions for the
spherical harmonic amplitudes of the potentials evidence that RWs
fulfil the trivial triadic resonance conditions
\begin{equation}
  (i\pm j)m_0=im_0\pm jm_0,\quad \omega_{i\pm j}=\omega_i\pm
  \omega_j=(i\pm j)\omega.\label{eq:triv_tri}
\end{equation}

Usually RW give rise to MRW by means of Hopf-type bifurcations (see
for instance \cite{Ran82,SGN13,GNS16,GSGS19}). We now argue how
triadic resonances can be described in terms of RWs and MRWs. Close to
the bifurcation point a MRW can be approximated by $u_0+\epsilon
\delta u_0$ where $u_0$ is a RW and $\delta u_0$ the corresponding
perturbation (Floquet eigenmode, since RW are periodic orbits, see for
instance \cite{Kuz98,SaNe16,GaSt18}). For the sake of simplicity only
the azimuthal dependence is considered in the following and the
investigation about the role of the latitude is reserved to
\S\ \ref{sec:nlin_colat}. Consider a RW with azimuthal symmetry $m_0$
and rotation frequency $\omega$ having the energy mainly contained in
the $m_0$ mode so it can be approximated by
$u_o(t,\varphi)=u_{m_o}\text{e}^{im_0(\varphi-\omega t)}$. We
approximate now the perturbation as $\delta
u_0(t,\varphi)=\overline{u}_{m_1}(t)\text{e}^{im_1\varphi}=u_{m_1}(t)\text{e}^{i(m_1\varphi-\omega_1
  t)}$ where $\omega_1$ is the new frequency provided by the Hopf
bifurcation and $m_1$ is the azimuthal symmetry of the Floquet mode
(see \cite{GSGS19} for details). The quadratic terms $u_o\delta u_0$
appearing in the advection term of the Navier-Stokes equations will
then exite a resonant mode since
\begin{equation}
\text{e}^{im_0(\varphi-\omega
  t)}\cdot\text{e}^{i(m_1\varphi-\omega_1
  t)}=\text{e}^{i((m_0+m_1)\varphi-(m_0\omega+\omega_1)t)}.\label{eq:tri_res}
\end{equation}
The triadic resonance conditions will be then
\begin{eqnarray*}
  \omega_2&=&\omega_0+\omega_1, \quad\text{with}\quad \omega_0=m_0\omega\\
  m_2&=&m_0+m_1.
\end{eqnarray*}

We note that as MRW are quasiperiodic, the frequencies $\omega_0$ and
$\omega_1$ are generically incommensurable so MRW can be
mathematically expressed like solutions of the inertial equation as in
Eq. (A5a,b) of \citet{BTHW18}. The only condition is that the
axisymmetry of the base state needs to be broken by means of a Hopf
bifurcation (\cite{EZK92,CrKn91}). We also remark that if $m_1=m_0$
the nonlinear interactions of Eq. (\ref{eq:tri_res}) are added to the
self nonlinear interactions of $u_0$, so the trivial resonance
conditions given in Eq. (\ref{eq:triv_tri}) may prevail.

\section{Results}
\label{sec:res}

In this section we investigate triadic resonances involving different
spherical harmonic degree $m$ and order $l$ for a set of several RW
and MRW belonging to two different scenarios already studied in
\cite{GSGS19,GSGS20b}. The parameters selected for the study are an
aspect ratio $\chi=0.5$, a Reynolds number $\Ree=10^3$, and several
Hartmann numbers with $\Ha<12.2$. For these parameters the first
instability for the steady axisymmetric base flow is nonaxisymmetric
with $m=3$ azimuthal symmetry, and antisymmetric with respect to the
equator (\cite{TEO11}). This instability is usually called ``radial
jet'' since a strong jet flowing outwards from the inner sphere at
equatorial latitudes is developed (\cite{HJE06}).

The first bifurcated nonlinear RWs, either stable or unstable with
azimuthal symmetries $m \in {2, 3, 4}$, were computed by \cite{GaSt18}
where the dependence of the associated rotation frequencies $\omega$
on $Ha$ on was described. In the following we investigate resonances
among different modes that constitute a MRW. We consider several MRWs
that bifurcate from the different branches of RWs with azimuthal
symmetry $m=3$ and $m=4$ (see \cite{GaSt18}, \cite{GSGS19}, and
\cite{GSGS20b} for further details of these MRWs).

Two different bifurcation scenarios with increasing complexity and
different sequences of azimuthal symmetry breakings are selected to
strengthen the link between different types MRW and triadic
resonances. Because MRW are quasiperiodic flows (invariant tori) they
are classified according to the number (two, three or four) of
incommensurable frequencies as 2T, 3T or 4T (``T'' standing for
Tori). The first bifurcation scenario provides the usual description
of stable flows (RWs and MRWs) emerging from an axisymmetric base
state. In this case the sequence of azimuthal symmetries of the flow
is: $m=0$ (base state), $m=3$ (RWs), and $m=1$ (2T and 3T MRWs). For
the second scenario, a larger number of bifurcations are analyzed. In
this case the sequence of azimuthal symmetries is: $m=0$ (base state),
$m=4$ (unstable RWs and 2T MRW), $m=2$ (3T MRWs), and $m=1$ (4T MRWs
and chaotic flows). Considering this second scenario allows us to
demonstrate the occurrence of triadic resonances even for chaotic
flows.

Each solution has been obtained using direct numerical simulations of
Eqs. (\ref{eq:mom}-\ref{eq:div}), from which the initial transients
are discarded. The spherical harmonics series of the fields
(Eqs. (\ref{eq:serie_psi}) and (\ref{eq:serie_phi})) have been
truncated at an order and degree of $L_{\text{max}}=84$, and $N_r=40$
inner collocation points are used in the radial direction. The
computations have been already validated in \cite{GaSt18,GSGS20b} by
increasing $L_{\text{max}}$ and $N_r$ up to $126$ and $60$,
respectively. To detect resonant modes we analyse time series obtained
from the poloidal potential $\Phi$ since it is directly related with
the radial velocity component
\begin{align*}
  v_r&=-\frac{1}{r\sin\theta}\lp\sin\theta\partial^2_{\theta\theta}\Phi
  +\cos\theta\partial_\theta\Phi+\frac{1}{\sin\theta}\partial^2_{\varphi\varphi}\Phi \rp.
\end{align*}
The radial velocity component naturally reflects the structure of the
instabilities as it measures the jet outwards velocity at equatorial
latitudes and it is related with the meridional circulation developed
at larger latitudes.

\subsection{Identification of triadic resonances}
\label{sec:Iden_reson}

\begin{figure}[t!]
\hspace{0.mm}\includegraphics[width=0.95\linewidth]{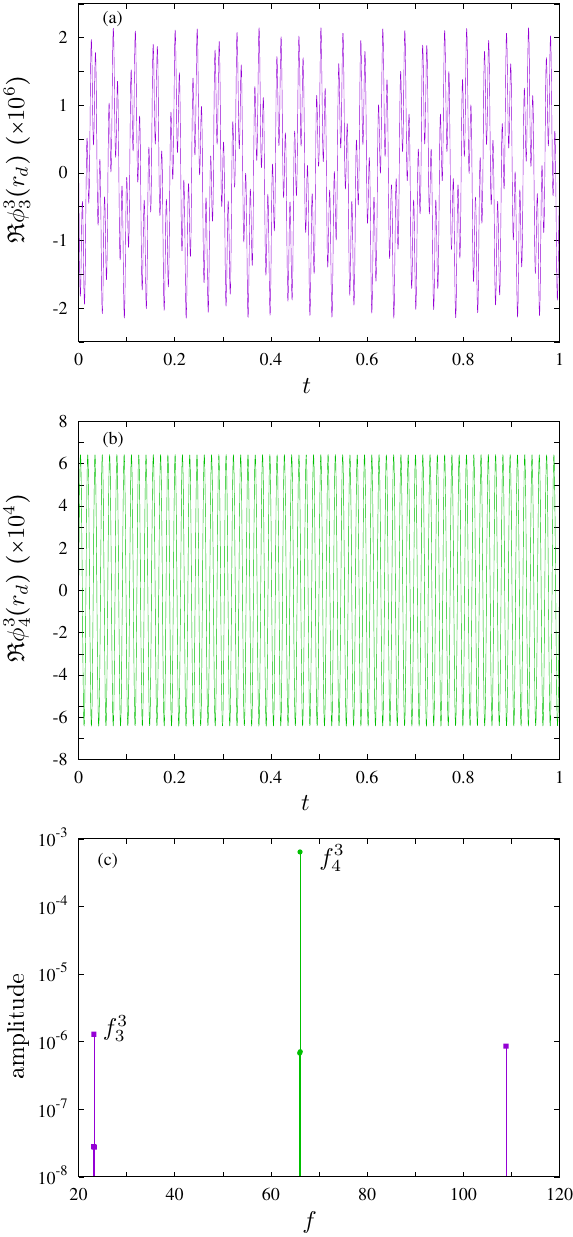}
\caption{(a) The time series of $\Re{\Phi^3_3(r_d)}$ is modulated. (b)
  The time series for $\Re{\Phi^3_4(r_d)}$ is periodic.  (c) The first
  four frequencies of largest amplitude from the frequency spectrum of
  $\Re{\Phi^3_3(r_d)}$ (squares) and of $\Re{\Phi^3_4(r_d)}$
  (circles). The peaks selected for the resonance relations shown in
  table \ref{table:res_2T_m1} are labelled on the plot. The time
  series in (a) and (b) is obtained for the 2T MRW solution at
  $\Ha=3.5$ which has $m=1$ azimuthal symmetry.}
\label{fig:m3_amp}
\end{figure}

For analyzing the frequency spectrum it is sufficient to consider
either the poloidal or the toroidal components, since both components
reflect the same frequencies in case of regular flows
(see e.g. Fig. 4a in \citet{BTHW18}).
Because of the symmetry, the time series of any equatorially symmetric
(resp. antisymmetric) poloidal mode, at a given radius, will exhibit the
same main peak in their frequency spectrum. That means that for each $m$
we incorporate the poloidal mode with $l = m$
(i. e. $\Phi_m^m$ which corresponds to an equatorially symmetric term)
and the poloidal mode with $l = m + 1$
(i.e. $\Phi_{m+1}^m$ which corresponds to an equatorially antisymmetric term)
\footnote{We recall that a poloidal term is equatorially symmetric if
  m + l is even and equatorially antisymmetric if m + l is odd.}.  In
order to identify triadic resonances, we specifically consider the
time series of the real part of the poloidal amplitude
$R\Phi_l^m(r_d)$, where $r_d = (r_i + r_o )/2$ is the middle shell
radius, for several azimuthal wave numbers $m \in {1, ..., 12}$ and
spherical harmonic degrees $l\in {1, ..., 13}$.

We compute the frequency spectrum of each time series by means of
Laskar’s algorithm (\citet{Las90,LFC92,Las93}), implemented in the
SDDSToolkit (\citet{BESS17}), which is an accurate method for the
determination of the fundamental frequencies of a time series up to
prescribed tolerance. We select the frequency among the four peaks
having largest amplitude in the way we now describe. The first peak is
selected when its amplitude is significantly larger than the
others. When there are several peaks of similar magnitude, we select
the frequency which provides the larger number of resonance
conditions, which is usually the largest peak. Figure \ref{fig:m3_amp}
exemplifies this procedure by displaying two examples of time series
and the corresponding frequency analysis for the modes
$(m,l)\in\{(3,3),(3,4)\}$. The time series originates from the 2T MRW
with azimuthal symmetry $m=1$ at $\Ha=3.5$ first studied in
\cite{GSGS19}. The time series of $R\Phi_3^3(r_d)$ (panel (a)) is
clearly quasiperiodic with the corresponding frequency spectrum
displaying two peaks of similar magnitude (panel (c), squares) whereas
the time series of $R\Phi_4^3(r_d)$ (panel (b)) is periodic with one
single strong peak in the spectrum (panel (c), circles).

When the frequency $f^m_l$ corresponding to each mode $(m,l)$ is
available, we seek for triadic resonances defined as
\begin{equation}
  f^{m_0}_{l_0}=f^{m_1}_{l_1}+f^{m_2}_{l_2},\quad  m_0=m_1+m_2,
\label{eq:res_rel}
\end{equation}  
for $m_i\in\{1,...,11\}$. The relative errors
$$\epsilon_f=(f^{m_0}_{l_0}-f^{m_1}_{l_1}-f^{m_2}_{l_2})/f^{m_0}_{l_0}$$
are computed for all the combinations satisfying $m_0=m_1+m_2$, and
$l_i=m_i$ or $l_i=m_i+1$ ($i=0,1,2$), with $m_0=3,....,12$. We assume
a resonance condition when the relative error is sufficiently small,
namely, $\epsilon_f<10^{-4}$. This value is similar to the accuracy
obtained for the determination of the fundamental frequency suggested
by the numerical experiments of \cite{GSGS21} performed in the same
parameter region as in the present study.

We note that there are resonances for which $l_0=l_1+l_2$ holds as
well, so the spherical harmonic degrees $l_i$ fulfill a similar
condition as the azimuthal wave numbers. In these cases the
latitudinal structure imposes an additional constraint for the
emergence of a triadic resonance. Note that for fixed $(m_i,l_i)$,
$i=0,1,2$, obeying the condition (\ref{eq:res_rel}), the condition
$f^{m_0}_{l_0}=f^{m_1}_{l_1+2k}+f^{m_2}_{l_2+2k}$, $k=1,2,..$ holds as
well since for modes $(m_i,l_i+2k)$ and $(m_i,l_i)$ have the same
frequency, as they have the same azimuthal wave number and equatorial
symmetry.  But even if the condition $l_0=l_1+l_2$ is not strictly
fulfilled, the resonance relations given in Eq.  (\ref{eq:res_rel})
have to respect the equatorial symmetry, in the sense that $l_0$ and
$l_1+l_2$ must have the same parity.

To check that the triad interactions preserve the same phase relation
during the whole time series a statistical measure, the bicoherence,
was estimated in \citet{BTHW18,HHT16}. In our study this is checked by
directly computing the phase $\xi^m_l$ associated to the frequencies
$f^m_l$ of the interacting modes $(m,l)$, which can be easily done
with Laskar's algorithm. The phase is computed on a specific time
window of length $T<T_f$, $T_f$ being the total length of the time
series, which is varied to obtain a time dependent value of the phase,
and also of the corresponding frequency (e.\,g. \cite{GSGS21}). For
each resonance relation like Eq. (\ref{eq:res_rel}) the corresponding
time dependent phases $\xi^m_l(t)$ must satisfy
\begin{equation}
  \xi^{m_0}_{l_0}(t)=\xi^{m_1}_{l_1}(t)+\xi^{m_2}_{l_2}(t),\quad  m_0=m_1+m_2,
\label{eq:res_ph}
\end{equation}  
up to a given tolerance.

\subsection{Path to triadic resonances from the base state}
\label{sec:s_wav}

\begin{figure}[t!]
\hspace{0.mm}\includegraphics[width=0.95\linewidth]{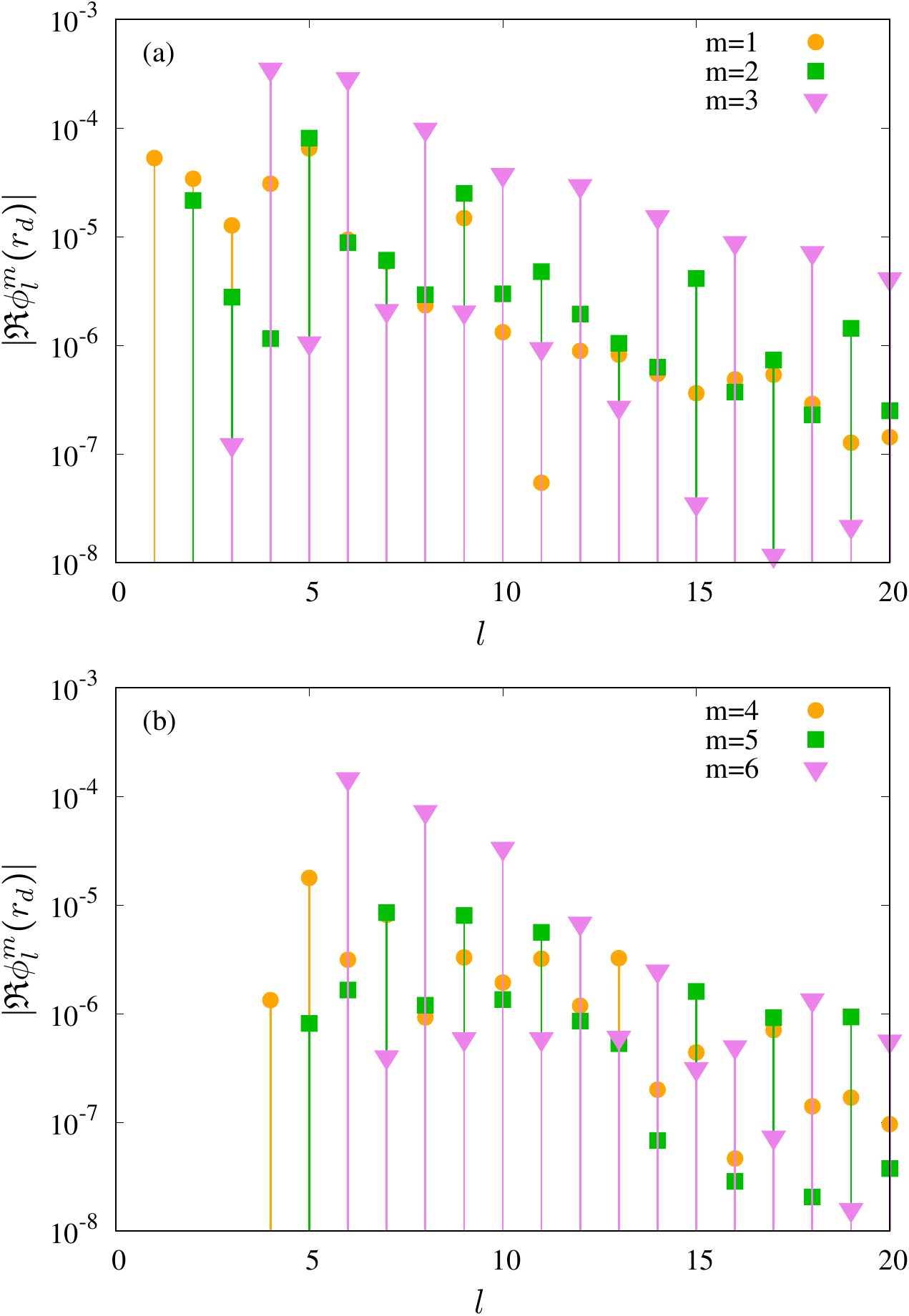}
\caption{Snapshot of the poloidal amplitudes $|\Re{\Phi^m_l(r_d)}|$,
  where $r_d=(r_i+r_o)/2$, versus the degree $l$ for
  $m\in\{1,..,6\}$. The solution is a modulated rotating wave with
  $m=1$ azimuthal symmetry at $\Ha=3.5$.}
\label{fig:Ha3.5_m3_ml}
\end{figure}

This subsection details the investigation of the development of
triadic resonances among spherical harmonic modes of nonlinear
solutions bifurcating directly from the steady axisymmetric base
state. At $\chi=0.5$ and $\Ree=10^3$ the onset of a nonaxisymmetric
radial jet instability is at $\Ha_c=12.2$, a value determined in
\cite{TEO11} from a linear stability analysis. The linear instability
takes the form of an equatorially antisymmetric radial jet with
azimuthal symmetry $m=3$ and thus equatorially asymmetric nonlinear
rotating waves with azimuthal symmetry $m=3$ are the first time
dependent flows (e.\,g. \cite{EZK92,GaSt18}). The spherical harmonic
amplitudes for the poloidal field (Eq.~(\ref{eq:serie_phi})) are given
by
\begin{eqnarray}
\Phi_{l}^{3k}(r,t)=\overline{\Phi}_{l}^{3k}(r)~e^{-i3k\omega t},\label{eq:RW3_phi}
\end{eqnarray}
with $\omega$ being the rotation frequency of the wave and
$\overline{\Phi}_{l}^{3k}(r)$ a function which only depends on the
radial coordinate (see \S\ \ref{sec:triad_reson}). We note that these
RW have in addition a more particular structure since
$\overline{\Phi}_{l}^{3k}(r)=0$ when $l$ is odd. This means that the
mode $m=3k$ is equatorially symmetric (resp. antisymmetric) if $k$ is
even (resp. odd). Because $f^m_l=m\omega/2\pi$ these RWs satisfy the
trivial triadic resonance conditions
$$f^3_4=f^{9}_{10}-f^6_6=f^{12}_{12}-f^{9}_{10}=f^{15}_{16}-f^{12}_{12}=...,$$
where the frequencies $f^m_l$ correspond to those of largest PSD of
the time series of $\Re{\Phi^m_l(r_d)}$. We note that for each
azimuthal mode $m=3k$, $k=1,2,...$, there is a latitudinal mode $l$
which verifies the resonance condition as well. For instance the
relation $f^3_4=f^{12}_{12}-f^{9}_{10}$ of above is equivalent to
$f^3_4=f^{12}_{14}-f^{9}_{10}$ since $f^{12}_{14}=f^{12}_{12}$.

\begin{table*}[t!]
\caption{Relations between the main frequencies $f^m_l$ of the
  different modes $(m,l)$ for the 2T MRW with $m=1$ azimuthal symmetry
  at $\Ha=3.5$ and the 3T MRW with $m=1$ azimuthal symmetry at
  $\Ha=3.10$. These relations are satisfied up to
  $(f^{m_0}_{l_0}-f^{m_1}_{l_1}-f^{m_2}_{l_2})/f^{m_0}_{l_0}<\epsilon_f$
  with $\epsilon_f=5\times 10^{-5}$. }
\label{table:res_2T_m1}
\renewcommand{\arraystretch}{1.5}
\begin{tabular}{|l|l|l|l|l|l|l|l|l|}
\hline
 $m=3$                     & $m=4$                       &  $m=5$                     & $m=6$                      & $m=7$                      & $m=8$                     & $m=9$                       & $m=10$                       & $m=11$                          \\
\hline
$f^3_3=f^1_1+f^2_2$         & $f^4_4=f^1_2+f^3_4$          & $f^5_5=f^1_1+f^4_4$         & $f^6_6=f^1_1+f^5_5$         & $f^7_7=f^1_1+f^6_6$          & $f^8_8=f^1_1+f^7_7$          & $f^9_9=f^1_1+f^8_8$         & $f^{10}_{10}=f^1_2+f^9_{10}$   & $f^{11}_{11}=f^1_1+f^{10}_{10}$    \\
                           & $\hspace{4.mm}=2f^2_3$      & $\hspace{4.mm}=f^1_2+f^4_5$ & $\hspace{4.mm}=f^1_2+f^5_6$ & $\hspace{4.mm}=f^2_2+f^5_5$ & $\hspace{4.mm}=f^2_2+f^6_6$ & $\hspace{4.mm}=f^2_2+f^7_7$ & $\hspace{5.mm}=f^2_3+f^8_9$  & $\hspace{5.mm}=f^1_2+f^{10}_{11}$ \\
                           &                             & $\hspace{4.mm}=f^2_3+f^3_4$ & $\hspace{4.mm}=f^2_2+f^4_4$ & $\hspace{4.mm}=f^2_3+f^5_6$ & $\hspace{4.mm}=f^2_3+f^6_7$ & $\hspace{4.mm}=f^3_3+f^6_6$ & $\hspace{5.mm}=f^3_4+f^7_8$  & $\hspace{5.mm}=f^2_3+f^{9}_{10}$  \\
                           &                             &                            & $\hspace{4.mm}=f^2_3+f^4_5$ & $\hspace{4.mm}=f^3_3+f^4_4$ & $\hspace{4.mm}=f^3_3+f^5_5$ & $\hspace{4.mm}=f^4_5+f^5_6$ & $\hspace{5.mm}=f^4_4+f^6_6$  & $\hspace{5.mm}=f^3_4+f^{8}_{9}$   \\
                           &                             &                            & $\hspace{4.mm}=2f^3_4$      & $\hspace{4.mm}=f^3_4+f^4_5$ & $\hspace{4.mm}=f^3_4+f^5_6$ &                            & $\hspace{5.mm}=f^4_5+f^6_7$  & $\hspace{5.mm}=f^4_4+f^7_7$      \\
                           &                             &                            &                             &                            & $\hspace{4.mm}=2f^4_5$      &                            & $\hspace{5.mm}=2f^5_5$      & $\hspace{5.mm}=f^4_5+f^7_8$      \\
                           &                             &                            &                             &                            &                            &                            &                             & $\hspace{5.mm}=f^5_5+f^6_6$      \\
                           &                             &                            &                             &                            &                            &                            &                             & $\hspace{5.mm}=f^5_6+f^6_7$      \\
\hline
$f^3_4=f^1_1+f^2_3$         & $f^4_5=f^1_1+f^3_4$          & $f^5_6=f^1_1+f^4_5$         & $f^6_7=f^1_2+f^5_5$         & $f^7_8=f^1_1+f^6_7$           & $f^8_9=f^1_1+f^7_8$         & $f^9_{10}=f^1_2+f^8_8$       & $f^{10}_{11}=f^1_1+f^9_{10}$  &$f^{11}_{12}=f^1_1+f^{10}_{11}$    \\
$\hspace{4.mm}=f^1_2+f^2_2$ & $\hspace{4.mm}=f^1_2+f^3_3$ & $\hspace{4.mm}=f^2_3+f^3_3$ & $\hspace{4.mm}=f^2_3+f^4_4$ & $\hspace{4.mm}=f^1_2+f^6_6$  & $\hspace{5.mm}=f^1_2+f^7_7$ & $\hspace{5.mm}=f^1_1+f^8_9$ & $\hspace{5.mm}=f^1_2+f^9_9$ & $\hspace{5.mm}=f^2_2+f^{9}_{10}$ \\
                           & $\hspace{4.mm}=f^2_2+f^2_3$ & $\hspace{4.mm}=f^2_2+f^3_4$ &                             & $\hspace{4.mm}=f^2_3+f^5_5$ & $\hspace{5.mm}=f^2_2+f^6_7$  & $\hspace{5.mm}=f^2_3+f^7_7$ & $\hspace{5.mm}=f^2_2+f^8_9$ & $\hspace{5.mm}=f^2_3+f^{9}_{9}$  \\
                           &                            &                             &                            & $\hspace{4.mm}=f^3_4+f^4_4$  & $\hspace{5.mm}=f^2_3+f^6_6$  & $\hspace{5.mm}=f^2_2+f^7_8$ & $\hspace{5.mm}=f^2_3+f^8_8$ & $\hspace{5.mm}=f^3_3+f^{8}_{9}$   \\
                           &                            &                             &                            &                             & $\hspace{5.mm}=f^3_4+f^5_5$  & $\hspace{5.mm}=f^3_4+f^6_6$ & $\hspace{5.mm}=f^3_3+f^7_8$ & $\hspace{5.mm}=f^3_4+f^8_8$      \\
                           &                            &                             &                            &                             & $\hspace{5.mm}=f^4_4+f^4_5$  & $\hspace{5.mm}=f^3_3+f^6_7$ & $\hspace{5.mm}=f^3_4+f^7_7$ & $\hspace{5.mm}=f^4_5+f^7_7$      \\
                           &                            &                             &                            &                             &                             & $\hspace{5.mm}=f^4_5+f^5_5$ & $\hspace{5.mm}=f^4_5+f^6_6$ & $\hspace{5.mm}=f^5_6+f^6_6$       \\
                           &                            &                             &                            &                             &                             & $\hspace{5.mm}=f^4_4+f^5_6$ & $\hspace{5.mm}=f^5_5+f^5_6$ &                                  \\
\hline
\end{tabular}
\end{table*}

According to \cite{GaSt18} the stability region of RWs with $m=3$
azimuthal symmetry extends down to $\Ha=3.95$ and a secondary Hopf
bifurcation gives rise to quasiperiodic 2T MRWs with azimuthal
symmetry $m=1$. For these types of solutions several resonance
conditions, reflecting the nonlinear interactions among the modes, are
found. To provide an idea of the contributions of the different modes
for a selected MRW at $\Ha=3.5$, figure \ref{fig:Ha3.5_m3_ml} displays
the poloidal amplitudes $|\Re{\Phi^m_l(r_d)}|$ at a fixed time instant
for different values of $m$ and $l$ (see figure caption). In contrast
to the special structure of RWs, where the equatorial symmetry of the
mode $m$ depends on its parity, this figure demonstrates that each
mode $m$ is equatorially asymmetric, as all the degree $l\ge m$ are
nonzero. However, the main structure of the RWs still prevails as the
dominant mode $m=3$ has a significantly larger equatorially
antisymmetric component and for the $m=6$ mode the largest component
is the equatorially symmetric. The $m=2,4,5$ modes have a dominant
equatorially antisymmetric ($m=2,4$) or symmetric ($m=5$) component
but both components are of similar magnitude in the case of $m=1$. The
patterns of the radial velocity for the different components of this
solution were analyzed in \cite{GSGS19} (concretely Figs. 6 and 7)
where the interested reader will find a comprehensive description.

Following the procedure described in \S\ \ref{sec:Iden_reson} we have
identified a multitude of triadic resonance relations among different
modes of the 2T MRW at $\Ha=3.5$. The relations are listed in table
\ref{table:res_2T_m1}. The first row of this table displays the
resonance relations $f^{m_0}_{m_0}=f^{m_1}_{l_1}+f^{m_2}_{l_2}$, for
each $m_0\in\{3,...,11\}$ from left to right, corresponding to the
equatorially symmetric mode $(m,l)=(m_0,m_0)$. The second row displays
$f^{m_0}_{m_0+1}=f^{m_1}_{l_1}+f^{m_2}_{l_2}$, which corresponds to
the resonance relation for the equatorially antisymmetric mode
$(m,l)=(m_0,m_0+1)$.

We note that for all the frequencies $f^{m_0}_{m_0}$ of the
equatorially symmetric modes, with the exception of $m_0=4,10$, a
resonance relation satisfying
$f^{m_0}_{m_0}=f^{1}_{1}+f^{m_0-1}_{m_0-1}$ can be found, which means
that both the azimuthal and latitudinal wave numbers are resonant. For
$m_0=4,10$ the resonance relation involving azimuthal and latitudinal
wave numbers is extracted from those of $m_0=5,11$, i.\,e.,
$f^4_4=f^5_5-f^1_1$ and $f^{10}_{10}=f^{11}_{11}-f^1_1$. The same
happens for the antisymmetric modes where either the relations
$f^{m_0}_{m_0+1}=f^{1}_{1}+f^{m_0-1}_{m_0}$ or
$f^{m_0}_{m_0+1}=f^{1}_{2}+f^{m_0-1}_{m_0-1}$ can be found.

It is interesting to remark that 2T MRWs with $m=1$ azimuthal symmetry
arise from RWs with $m=3$ symmetry by applying perturbations (Floquet
modes, already computed in \cite{GaSt18}) of azimuthal symmetry
$m=1$. Then, the nonlinear coupling (described in
\S\ \ref{sec:triad_reson}) is responsible for the resonant
interactions described for the 2T MRW at $\Ha=3.5$ for which the modes
$(m,l)=\{(1,1),(1,2)\}$ are in resonance with all the other modes
investigated in Table \ref{table:res_2T_m1}. The triadic resonance
relations found for the 2T MRW at $\Ha=3.5$ have been found also for
2T MRWs belonging to the same branch which is stable for
$\Ha\in[3.4,3.94]$. In addition, we have also found that the resonance
relations of Table \ref{table:res_2T_m1} are conserved for 3T MRWs
that bifurcates from the 2T MRW at $\Ha=3.395$ and which are stable
down to $\Ha=3.1$. Then, the resonance interactions take place for all
solutions obtained in the range $\Ha\in[3.4,3.94]$, which is the range
of stability of periodic and quasiperiodic flows which bifurcate from
the base state.

\begin{figure}
\hspace{0.mm}\includegraphics[width=0.93\linewidth]{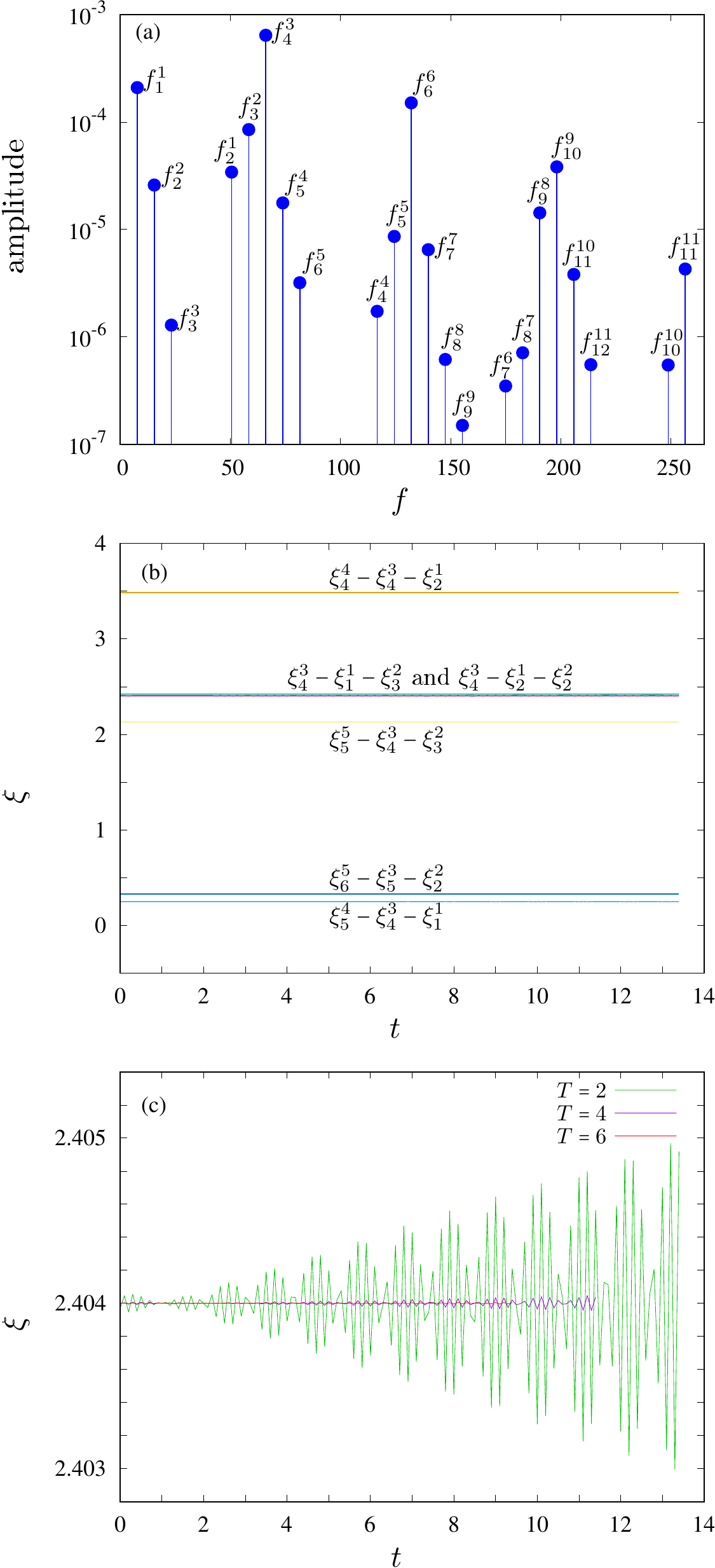}
\caption{Analysis for a 2T MRW with $m=1$ azimuthal symmetry at
  $\Ha=3.5$.  (a) Leading frequencies and amplitudes of the time
  series of poloidal component of the different modes $(m,l)$. The
  frequency $f^m_l$ corresponds to the main peak of the frequency
  spectrum of the time series of $\Re{\Phi^m_l(r_d)}$. (b) Phase
  coupling among the different modes corresponding to different types
  of resonances shown in Table \ref{table:res_2T_m1}. The phase of
  each frequency $f^m_l$ is $\xi^m_l$. (c) The coupling
  $\xi^3_4=\xi^1_1+\xi^2_3$ is investigated by a time-dependent
  frequency analysis with time windows of length $T$.}
\label{fig:Ha3.5_m3_freq}
\end{figure}

Figure \ref{fig:Ha3.5_m3_freq}(a) shows the frequencies $f^m_l$
involved in Table \ref{table:res_2T_m1} versus their corresponding
amplitudes. The relations $f^{m_0}_{m_0}<f^{m_0+1}_{m_0+1}$ and
$f^{m_0}_{m_0+1}<f^{m_0+1}_{m_0+2}$ are verified for all $m_0=1,..,11$
meaning that low wave numbers have smaller frequencies (slow modes)
and larger wave numbers have larger frequencies (fast modes). The
figure shows however that no clear threshold between slow and fast
modes can be defined. In contrast to the results in \citet{BTHW18},
where only resonances between one slow and two fast modes are
reported, we find resonances between slow modes
(e.\,g. $f^3_3=f^1_1+f^2_2$), fast and slow modes
(e.\,g. $f^{11}_{11}=f^1_1+f^{10}_{10}$), and fast modes
(e.\,g. $f^{10}_{10}=f^4_4+f^6_6$).

Figures \ref{fig:Ha3.5_m3_freq}(b) and (c) correspond to the analysis
of the phase coupling among the resonant modes, described in
\S\ \ref{sec:Iden_reson}. Concretely, the time dependent relations
given in Eq. (\ref{eq:res_ph}) are displayed in
Fig. \ref{fig:Ha3.5_m3_freq}(b) for some resonant triads. The time
window used for this figure is $T=2$ over a time series covering the
interval $[0,15]$. The phase coupling is satisfied up to an accuracy
which is similar to that for the frequency determination as it is
demonstrated in Fig. \ref{fig:Ha3.5_m3_freq}(c). The latter figure
shows that the tiny variation of $\xi^3_4-\xi^1_1-\xi^2_3$ decreases
as the size of the time window $T$ increases, i.\,e. as the accuracy
of the frequency determination increases.

\subsection{Path to triadic resonances from unstable rotating waves}
\label{sec:uns_wav}

Unstable RWs with azimuthal symmetry $m=4$ bifurcate from the
axisymmetric base state at a Hartmann number $\Ha=11.04<\Ha_c=12.2$
(see \cite{GaSt18}), where $\Ha_c$ is the critical value predicted by
the linear theory (\cite{TEO11}). These RWs again correspond to the
radial jet instability. Because the linear eigenfunctions are
equatorially antisymmetric, and RWs are nonlinear solutions, these RWs
are equatorially asymmetric. They have however a particular structure
(as happened for RWs described in \S\ \ref{sec:s_wav}) since the
$m=4k$ modes are equatorially antisymmetric (resp. symmetric) when $k$
is an odd (resp. even) integer number. Then, if $\omega$ is the
rotation frequency of the RW the spherical harmonic amplitudes for the
poloidal field of Eq.~(\ref{eq:serie_phi}) read
\begin{eqnarray}
\Phi_{l}^{4k}(r,t)=\overline{\Phi}_{l}^{4k}(r)~e^{-i4k\omega t},\label{eq:RW4_phi}
\end{eqnarray}
with $\overline{\Phi}_{l}^{4k}(r)=0$ when $k+l$ is an odd integer
number.

A secondary Hopf bifurcation (see \cite{GSGS20b} gives rise to 2T MRWs
with the same azimuthal symmetry and the same mode structure. In this
case the spherical harmonic amplitudes $\Phi_{l}^{4k}(r,t)$ become
quasiperiodic of the form 
\begin{eqnarray}
\Phi_{l}^{4k}(r,t)=\overline{\Phi}_{l}^{4k}(t,r)~e^{-i4k\omega t},\label{eq:MRW4_phi}
\end{eqnarray}
with $\overline{\Phi}_{l}^{4k}(t,r)$ being a periodic function with
$\overline{\Phi}_{l}^{4k}(t,r)=0$ when $k+l$ is an odd integer number,
and $\omega$ the rotation frequency of the MRW (e.\,g. \cite{Ran82}).

Because of these structures, both RWs and MRWs, fulfill the triadic
resonance conditions for the azimuthal wave number $m$:
\begin{equation}
  f^4_5=f^{12}_{13}-f^8_8=f^{16}_{16}-f^{12}_{13}=f^{20}_{21}-f^{16}_{16}=...,
\label{eq:res_2T}
\end{equation}
where the frequencies $f^m_l$ correspond to those of largest amplitude
in the frequency spectrum of the time series of
$\Re{\Phi^m_l(r_d)}$. As in the previous section, for each azimuthal
mode $m=4k$, $k=1,2,...$, there is a latitudinal mode $l$ which verify
the resonance condition. For instance the relation
$f^4_5=f^{16}_{16}-f^{12}_{13}$ given above corresponds to
$f^4_5=f^{16}_{18}-f^{12}_{13}$, since $f^{16}_{18}=f^{16}_{16}$.

By further varying the Hartmann number a tertiary Hopf bifurcation
gives rise to 3T MRWs, i.\,e. quasiperiodic flows with three
incommensurable frequencies. By means of this bifurcation the $m=4$
azimuthal symmetry is broken and the new MRWs acquire azimuthal
symmetry $m=2$. In contrast to MRWs with $m=4$ azimuthal symmetry, all
the azimuthal modes $m$ are now equatorially asymmetric, meaning that
$\Phi^m_m(t,r),\Phi^{m}_{m+1}(t,r),....,\Phi^{m}_{L_{\text{max}}}(t,r)$
are nonzero. The dominant azimuthal wave numbers for this solution are
$m=4,2,8,6,10,12$ (see Fig.~(5) of \cite{GSGS20b}) so we focus on
studying the resonances among these modes. We have identified several
relations between the frequencies $f^m_l$, which are listed in Table
\ref{table:res_3T_m2}. These relations include those of
Eq.~(\ref{eq:res_2T}) found for the parent RWs and MRWs. The fact that
bifurcated solutions conserve the resonance relations was also found
in the scenario studied in \S\ \ref{sec:s_wav}. As will be evidenced
in the following this is indeed true for further bifurcations found in
\cite{GSGS20b}.

\begin{table*}[t!]
\caption{Relations between the main frequencies $f^m_l$ of the
  different modes $(m,l)$ for the 3T MRW with $m=2$ azimuthal symmetry
  at $\Ha=1.6$.}
\label{table:res_3T_m2}
\renewcommand{\arraystretch}{1.5}
\begin{tabular}{|l|l|l|l|l|}
\hline
$m=4$ & $m=6$ & $m=8$ & $m=10$ & $m=12$ \\
\hline
$f^4_4=2f^2_3$      & $f^6_6=f^2_2+f^4_4=f^2_3+f^4_5$ & $f^8_8=f^2_2+f^6_6=f^2_3+f^6_7=2f^4_5$ & $f^{10}_{10}=f^2_2+f^8_8=f^4_5+f^6_7$ &$f^{12}_{12}=f^2_2+f^{10}_{10}=2f^6_7$\\
\hline
$f^4_5=f^2_2+f^2_3$ & $f^6_7=f^2_2+f^4_5$             & $f^8_9=f^2_3+f^6_6=f^4_4+f^4_5$  & $f^{10}_{11}=f^2_2+f^8_9=f^2_3+f^8_8$&$f^{12}_{13}=f^2_2+f^{10}_{11}=f^2_3+f^{10}_{10}$\\
                   &                                &                                & $\hspace{5.mm}=f^4_4+f^6_7=f^4_5+f^6_6$          & $\hspace{5.mm}=f^4_5+f^8_8=f^6_6+f^6_7$ \\
\hline
\end{tabular}
\end{table*}

\begin{figure}
\hspace{0.mm}\includegraphics[width=0.95\linewidth]{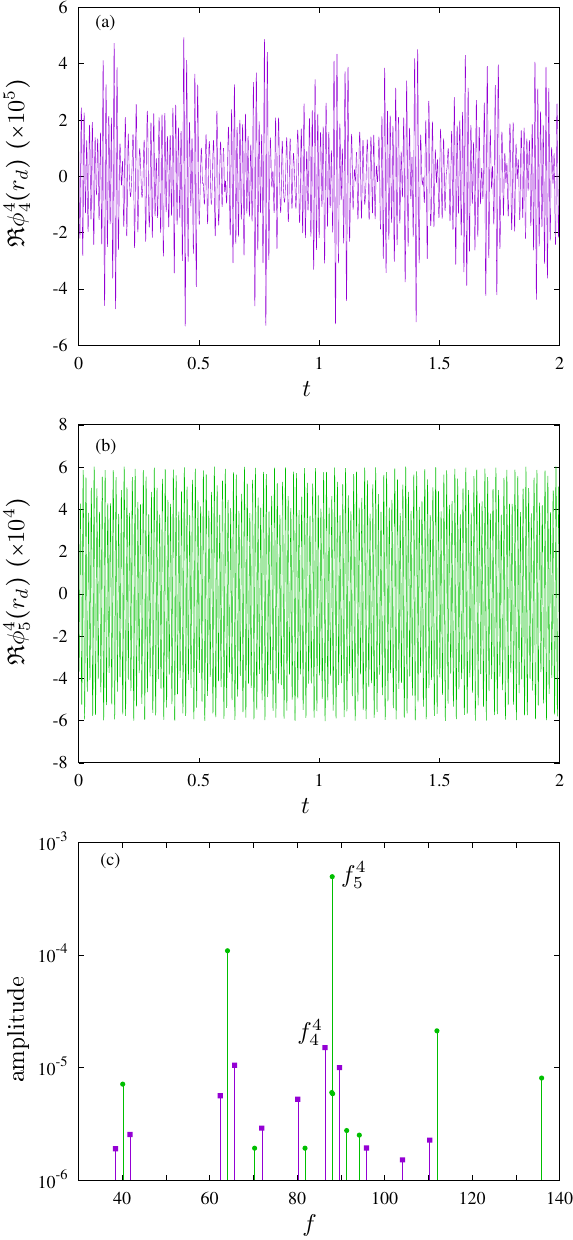}
\caption{(a) The time series of $\Re{\Phi^4_4(r_d)}$ is strongly
  modulated. (b) The time series for $\Re{\Phi^4_5(r_d)}$ is slightly
  modulated.  (c) The first ten frequencies of largest amplitude from
  the frequency spectrum of $\Re{\Phi^4_4(r_d)}$ (squares) and of
  $\Re{\Phi^4_5(r_d)}$ (circles). The peaks selected for the resonance
  relations shown in table \ref{table:res_4TC_m1} are labelled on the
  plot. The solution is a 4T MRW with $m=1$ azimuthal symmetry at
  $\Ha=1.4$.}
\label{fig:m4_amp}
\end{figure}

After a quaternary Hopf bifurcation, totally breaking the azimuthal
symmetry, quasiperiodic MRWs with four incommensurable frequencies
develop and the resonance relations for the same 4T MRW analyzed in
\cite{GSGS20b}, at $\Ha=1.4$, are provided in Table
\ref{table:res_4TC_m1}. There are less relations when comparing with
Table \ref{table:res_2T_m1} of \S\ \ref{sec:s_wav}, obtained for a 2T
MRW, which seems reasonable since the time dependence is now more
complex as illustrated in figure \ref{fig:m4_amp}(a,b). The time
series of $\Re{\Phi^m_l(r_d)}$ have now several comparable peaks in
their frequency spectrum so the selection of the frequency $f^m_l$
becomes less clear (compare Fig. \ref{fig:m3_amp}(c) with
Fig. \ref{fig:m4_amp}(c)).

\begin{table*}[ht]
\caption{Relations between the main frequencies $f^m_l$ of the
  different modes $(m,l)$ for the 4T MRW with $m=1$ azimuthal symmetry
  at $\Ha=1.4$ and a chaotic flow with $m=1$ azimuthal symmetry at
  $\Ha=0.5$. For $m=3$ only the relation $f^3_3=f^1_1+f^2_2$ has been
  found. Note that these relations contain those for the 3T MRW with
  $m=2$ azimuthal symmetry shown in Table \ref{table:res_3T_m2}. These
  relations are satisfied up to
  $(f^{m_0}_{l_0}-f^{m_1}_{l_1}-f^{m_2}_{l_2})/f^{m_0}_{l_0}<\epsilon_f$
  with $\epsilon_f=4\times 10^{-5}$ for the 4T MRW and with
  $\epsilon_f=1.5\times 10^{-4}$ for the chaotic flow.}
\label{table:res_4TC_m1}
\renewcommand{\arraystretch}{1.5}
\begin{tabular}{|l|l|l|l|l|l|l|l|l|}
\hline
 $m=4$                     &  $m=5$             & $m=6$                      & $m=7$                      & $m=8$                     & $m=9$                       & $m=10$                       & $m=11$                          & $m=12$ \\
\hline
$f^4_4=2f^2_3$              & $f^5_5=f^2_3+f^3_4$ & $f^6_6=f^1_1+f^5_5$         & $f^7_7=f^2_2+f^5_5$         & $f^8_8=f^1_1+f^7_7$          & $f^9_9=f^2_2+f^7_7$         & $f^{10}_{10}=f^1_1+f^9_9$     & $f^{11}_{11}=f^1_1+f^{10}_{10}$     & $f^{12}_{12}=f^2_2+f^{10}_{10}$  \\
                           &                    & $\hspace{4.mm}=f^2_2+f^4_4$  & $\hspace{4.mm}=f^2_3+f^5_6$ & $\hspace{4.mm}=f^2_2+f^6_6$ & $\hspace{4.mm}=f^3_4+f^6_7$ & $\hspace{5.mm}=f^2_2+f^8_8$ & $\hspace{5.mm}=f^3_3+f^{8}_{8}$   & $\hspace{5.mm}=f^3_3+f^{9}_{9}$ \\
                           &                    & $\hspace{4.mm}=f^2_3+f^4_5$ & $\hspace{4.mm}=f^3_4+f^4_5$ & $\hspace{4.mm}=f^2_3+f^6_7$ & $\hspace{4.mm}=f^4_5+f^5_6$ & $\hspace{5.mm}=f^3_3+f^7_7$  & $\hspace{5.mm}=f^4_5+f^{7}_{8}$  & $\hspace{5.mm}=f^5_6+f^{7}_{8}$ \\
                           &                    &                            &                            & $\hspace{4.mm}=f^3_3+f^5_5$ &                             & $\hspace{5.mm}=f^3_4+f^7_8$ &                                 & $\hspace{5.mm}=2f^{6}_{7}$      \\
                           &                    &                            &                            & $\hspace{4.mm}=2f^4_5$      &                             & $\hspace{5.mm}=f^4_5+f^6_7$ &                                 &                                \\
\hline
$f^4_5=f^1_1+f^3_4$         & $f^5_6=f^2_2+f^3_4$ & $f^6_7=f^1_1+f^5_6$         & $f^7_8=f^1_1+f^6_7$         & $f^8_9=f^2_3+f^6_6$          & $f^9_{10}=f^2_3+f^7_7$         & $f^{10}_{11}=f^1_1+f^9_{10}$  &$f^{11}_{12}=f^2_2+f^{9}_{10}$   &$f^{12}_{13}=f^1_1+f^{11}_{12}$            \\
$\hspace{4.mm}=f^2_2+f^2_3$ &                    & $\hspace{4.mm}=f^2_2+f^4_5$ & $\hspace{4.mm}=f^3_3+f^4_5$ & $\hspace{4.mm}=f^4_4+f^4_5$ & $\hspace{5.mm}=f^3_4+f^6_6$  & $\hspace{5.mm}=f^2_2+f^8_9$ & $\hspace{5.mm}=f^2_3+f^{9}_{9}$ & $\hspace{5.mm}=f^2_2+f^{10}_{11}$ \\
                           &                    & $\hspace{4.mm}=f^3_3+f^3_4$ &                            &                             & $\hspace{5.mm}=f^4_5+f^5_5$ & $\hspace{5.mm}=f^2_3+f^8_8$ & $\hspace{5.mm}=f^3_4+f^{8}_{8}$  & $\hspace{5.mm}=f^2_3+f^{10}_{10}$ \\
                            &                   &                            &                            &                             & $\hspace{5.mm}=f^4_4+f^5_6$ & $\hspace{5.mm}=f^4_4+f^6_7$ & $\hspace{5.mm}=f^4_5+f^{7}_{7}$  & $\hspace{5.mm}=f^3_3+f^{9}_{10}$ \\
                            &                   &                            &                            &                             &                            & $\hspace{5.mm}=f^4_5+f^6_6$ & $\hspace{5.mm}=f^5_5+f^6_7$     & $\hspace{5.mm}=f^4_5+f^8_8$ \\
                            &                   &                            &                            &                             &                            &                            & $\hspace{5.mm}=f^5_6+f^6_6$     & $\hspace{5.mm}=f^5_5+f^7_8$ \\
                            &                   &                            &                            &                             &                            &                            &                                 & $\hspace{5.mm}=f^6_6+f^6_7$ \\
\hline
\end{tabular}
\end{table*}

\begin{figure}
\hspace{0.mm}\includegraphics[width=0.95\linewidth]{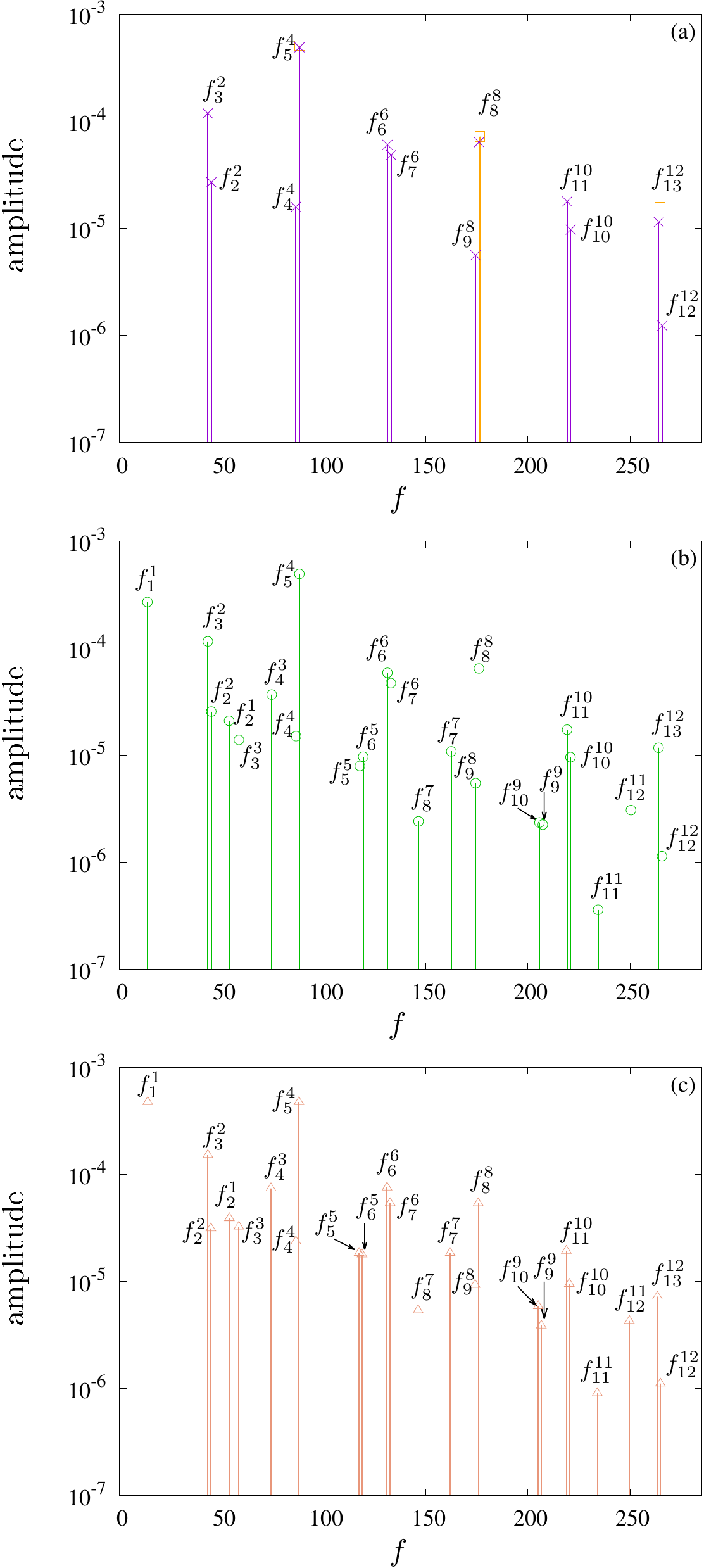}
\caption{Leading frequencies and amplitudes of the time series of
  poloidal component of the different modes $(m,l)$. The frequency
  $f^m_l$ corresponds to one peak of the frequency spectrum of the
  time series of $\Re{\Phi^m_l(r_d)}$, where $r_d=(r_i+r_o)/2$. (a)
  The solutions are a 2T MRW with $m=4$ azimuthal symmetry
  (squares) and a 3T MRW with $m=2$ azimuthal symmetry (crosses),
  both at $\Ha=1.6$. (b) The solution is a 4T MRW with $m=1$ azimuthal
  symmetry (circles) at $\Ha=1.4$. (c) The solution is a chaotic wave
  with $m=1$ azimuthal symmetry (triangles) at $\Ha=0.5$.}
\label{fig:m4_freq}
\end{figure}

The resonance relations found for the 3T MRWs prevail for the
bifurcated 4T MRWs (all relations of Table \ref{table:res_3T_m2} are
included in Table \ref{table:res_4TC_m1}). Moreover, the resonance
relations for the 4T MRWs of Table \ref{table:res_4TC_m1} even prevail
for chaotic flows bifurcated from these 4T MRWs. We have checked this
by considering a chaotic flow at $\Ha=0.5$. As noticed in
\cite{GSGS20,GSGS20b} the main frequencies associated to the flow and
to the volume-averaged properties remain nearly constant for a wide
interval of Hartmann numbers, at which several types of 2T, 3T and 4T
MRW can be found. Figure \ref{fig:m4_freq} evidences that this is also
true for the frequencies $f^m_l$. Figure \ref{fig:m4_freq} shows from
top to bottom the selected frequencies $f^m_l$ and amplitudes for the
sequence of bifurcated solutions: 2T MRW and 3T MRW both at $\Ha=1.6$
(panel (a)), 4T MRW at $\Ha=1.4$ (panel (b)), and a chaotic flow at
$\Ha=0.5$ (panel (c)). From this figure it is clear that $f^m_l$ are
almost unchanged.

\begin{figure*}
\hspace{0.mm}\includegraphics[width=0.95\linewidth]{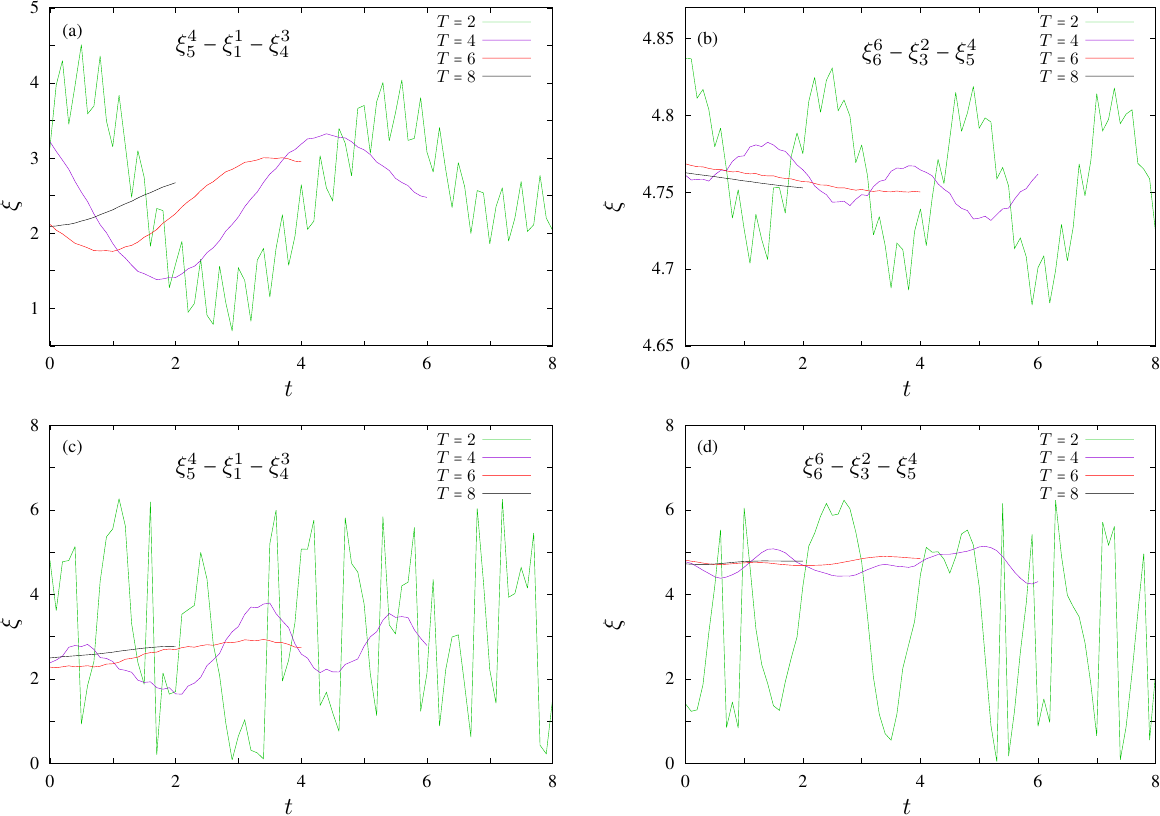}
\caption{Phase coupling among the different modes corresponding to two
  different types of resonances $f^4_5=f^1_1+f^3_4$ and
  $f^6_6=f^2_3+f^4_5$ shown in Table \ref{table:res_4TC_m1}. The phase
  of each frequency $f^m_l$ is $\xi^m_l$. The coupling is investigated
  by a time-dependent frequency analysis with time windows of length
  $T$. (a,c) The relation $\xi^4_5(t)-\xi^1_1(t)-\xi^3_4(t)$ tends to
  be roughly constant as the length of the time window $T$
  increases. (b,d) The relation $\xi^6_6(t)-\xi^2_3(t)-\xi^4_5(t)$ is
  nearly constant as the length of the time window $T$
  increases. (a,b) Correspond to a 4T MRW at $\Ha=1.4$ and (c,d)
  correspond to a chaotic wave at $\Ha=0.5$.}
\label{fig:m4_phase}
\end{figure*}

Figure \ref{fig:m4_phase} is concerned with the analysis of the phase
coupling among the resonant modes (see \S\ \ref{sec:Iden_reson}) for
the 4T MRW (top row) at $\Ha=1.4$ and for the chaotic flow at
$\Ha=0.5$ (bottom row). In this bifurcation scenario the time
dependence of the phase coupling relations
$\xi^{m_0}_{l_0}-\xi^{m_1}_{l_1}-\xi^{m_2}_{l_2}$, although
noticeable, is being damped as the time window $T$ increases. This
means that for sufficiently large time windows the phase coupling
tends to be verified. For the 4T and chaotic flow, the necessity of
considering very large time windows ($T>10$) in time dependent
frequency analysis has been already pointed out in \cite{GSGS20b}.

\subsection{Comparison with triadic resonances of inertial modes}
\label{sec:comp}

We now compare quantitatively our results with the previous studies of
\citet{BTHW18} and \citet{Lin21} in rotating spherical geometry for
which triadic resonances, involving inertial modes, are described.  In
the first place we compare with the experiments and simulations of
\citet{BTHW18} as the analyzed problem, the SC flow with a rotating
inner (with $\Omega_i$) and outer (with $\Omega_o$) spheres, is a
mechanically driven system equivalent to the problem considered here,
the MSC flow. In the second place, we compare with the thermal
convection problem in a full sphere of \citet{Lin21}. Although we are
contrasting problems with different types of driving or dynamical
regimes the comparison makes sense since all the systems are
{\bf{SO}}$(2)\times${\bf{Z}}$_2$-equivariant.

The frequencies shown in Table \ref{table:res_2T_m1} and
fig. \ref{fig:Ha3.5_m3_freq}(a) are dimensionless. With the time scale
$d^2/\nu$ considered in our modelling, dimensional frequencies, $f_d$, are
obtained by $f_d=f\nu/d^2$ so
\begin{equation*}
\frac{\omega_d}{\Omega_i}=\frac{2\pi f_d}{\Omega_i}=\frac{\omega}{\Ree(\chi-1)},
\end{equation*}
with $\omega=2\pi f$. The ratios $\omega_d/\Omega_i$ are the angular
frequences normalized by the angular frequency of rotation of the
inner sphere since the outer sphere is at rest, $\Omega_o=0$. For the
modes analyzed in fig. \ref{fig:Ha3.5_m3_freq}(a) the slowest mode is
$(m,l)=(1,1)$ with a ratio $\omega_d/\Omega_i=0.097$. This value is
very similar to that of the basic $m=1$ instability studied in
\citet{BTHW18} with a ratio $\omega_d/\Omega_o=0.086$. In our case the
fastest mode corresponds to $(m,l)=(11,11)$ with a ratio
$\omega_d/\Omega_i=3.22$. The mode $(m,l)=(3,4)$, which has the
largest peak in the frequency spectrum, has a ratio close to unity
$\omega_d/\Omega_i=0.83$. In the case of the SC flow (\citet{BTHW18})
values in the range $(0.4,1)$ have been obtained for the ratio
$\omega_d/\Omega_o$ in the case of fast inertial modes.

The development of triadic resonances for rotating thermal convection
in a full sphere studied by \cite{Lin21} follows the same picture as
investigated here. Rotating waves in the form of inertial modes with
$m=4$ (e.\,g. \cite{Zha93,NGS08,SGN16b}) are the first time-dependent
instabilities in the considered parameter regime (low Prandtl and
Ekman numbers). \cite{Lin21} slightly increases the forcing parameter
(the Rayleigh number) to obtain an azimuthally asymmetric
quasiperiodic flow with two frequencies (see their Fig. 1), i.\,e. a
2T MRW with azimuthal symmetry $m=1$, for which resonant azimuthal
modes have been found. The appearance of MRWs in rotating thermal
convection in spherical shells due to a Hopf bifurcation from RWs is a
generic scenario which has been studied in \cite{GSDN15,GNS16}. In
addition, and in agreement with our results, \cite{Lin21} has also
found resonant behaviour for chaotic flows close to the onset.

For the inertial convective resonant solutions the value
$\omega_d/\Omega$ has been also computed (see Table 1 of \cite{Lin21})
to identify fast and slow modes. By varying the Ekman and Prandtl
numbers it was found that $|\omega_d/\Omega|\in(0.0899,0.5352$, either
for the primary or free modes, which are $m=\{4,1,3\}$, $m=\{1,7,8\}$,
and $m=\{3,10,13\}$ for the three different cases studied in Table 1
of \cite{Lin21}. These results are indeed similar to the results for
the radial jet instability (Secs. \ref{sec:s_wav} and
\ref{sec:uns_wav}) since resonances involving these azimuthal wave
numbers are found with values $\omega_d/\Omega_i$ of similar
magnitude.

\section{Nonlinear interactions among latitudinal modes}
\label{sec:nlin_colat}

\begin{figure}
  \hspace{0.mm}\includegraphics[width=0.95\linewidth]{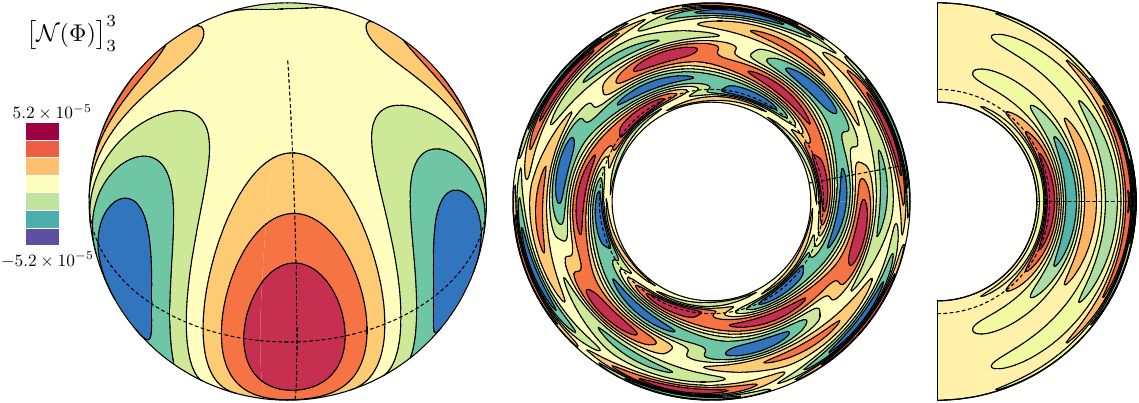}\\
  \hspace{0.mm}\includegraphics[width=0.95\linewidth]{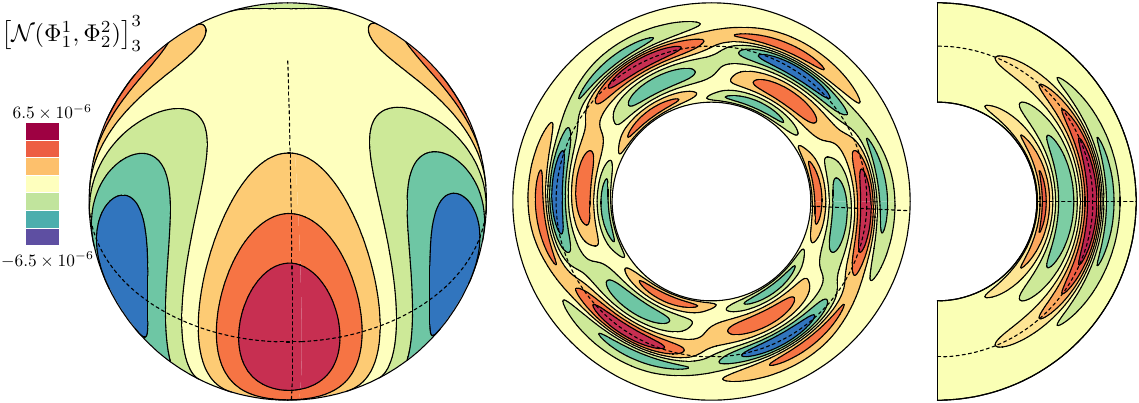}
\caption{Contour plots on spherical, latitudinal and meridional
  sections through a local maximum for the $(m,l)=(3,3)$ component of
  the nonlinear term corresponding to the Navier-Stokes equations for
  the poloidal field. The evaluation of this nonlinear term is applied
  to the poloidal field whereas the toroidal field is set to zero. All
  the spherical harmonics amplitudes are considered for the poloidal
  field in the top row but only the modes $(m,l)=(1,1),(2,2)$ are
  considered in the bottom row. The solution is a 2T MRW with $m=1$
  azimuthal symmetry at $\Ha=3.5$.}
\label{fig:ctplm3_m3l3}
\end{figure}

\begin{figure}
  \hspace{0.mm}\includegraphics[width=0.95\linewidth]{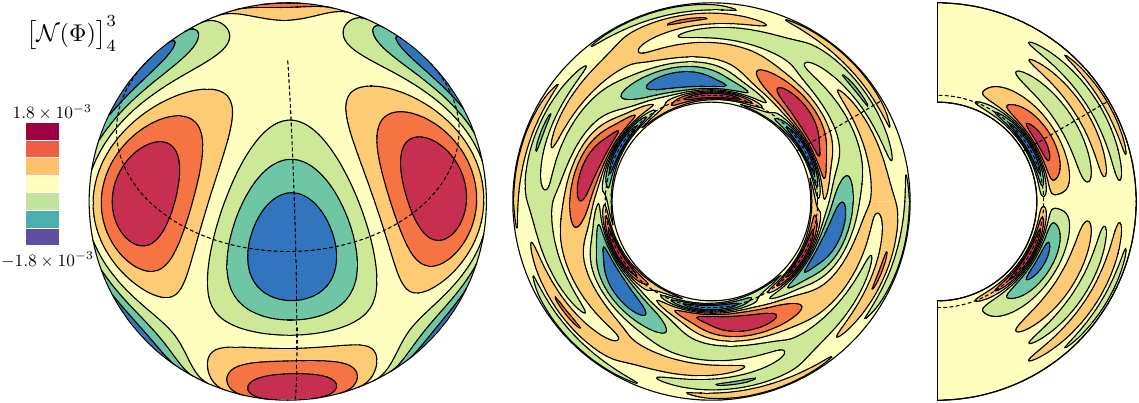}\\
  \hspace{0.mm}\includegraphics[width=0.95\linewidth]{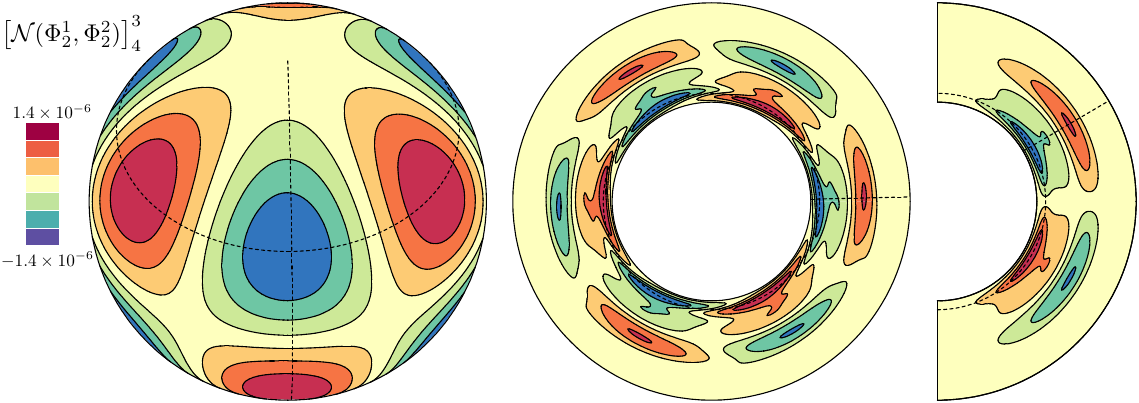}\\
  \hspace{0.mm}\includegraphics[width=0.95\linewidth]{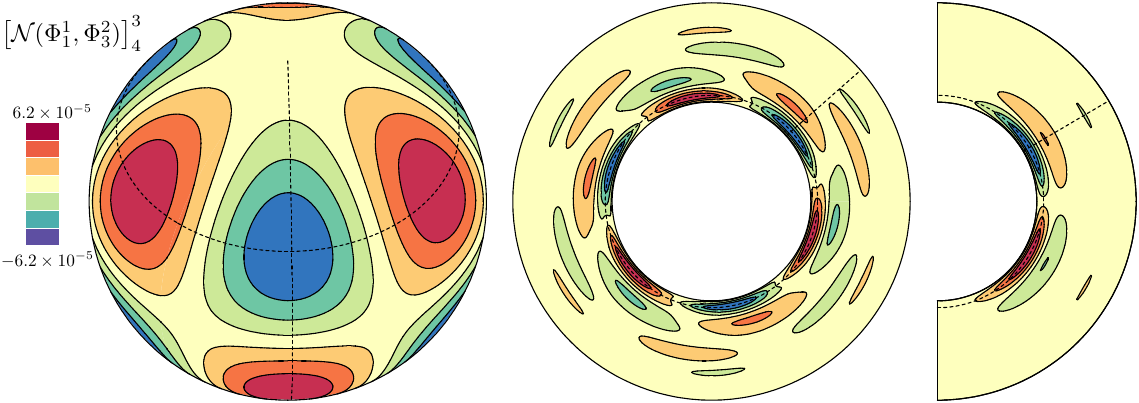}\\
\caption{As figure \ref{fig:ctplm3_m3l3} but for the $(m,l)=(3,4)$
  component of the nonlinear term. In this figure the modes
  $(m,l)=(1,2),(2,2)$ and $(m,l)=(1,1),(2,3)$ are considered in the
  middle and bottom row.  The solution is a 2T MRW with $m=1$
  azimuthal symmetry at $\Ha=3.5$.}
\label{fig:ctplm3_m3l4}
\end{figure}

The quadratic nonlinear nature of the advection term in the
Navier-Stokes equations provides a way to to understand the
development of the different resonances obtained in
\S\ \ref{sec:s_wav} and \S\ \ref{sec:uns_wav}. The azimuthal
coupling among the different resonant waves has been already described
in \S\ \ref{sec:triad_reson} and we now focus on the latitudinal
coordinate.

The product of the spherical harmonics expansion
(e.\,g. Eq.~(\ref{eq:serie_phi})) gives rise to products of associated
Legendre functions
$P^{m_1}_{l_1}(\cos{\theta})P^{m_2}_{l_2}(\cos{\theta})\equiv
P^{m_1}_{l_1}P^{m_2}_{l_2}$ which can be expressed in terms of a
linear combination of associated Legendre functions of order $m_1+m_2$
and appropriate degree $l$ (see \cite{Hwa95}):
\begin{equation}
  P^{m_1}_{l_1}P^{m_2}_{l_2}=\sum_{l=0}^{\big [\nu/2 \big ]}D_{l_1+l_2-2l}P^{m_1+m_2}_{l_1+l_2-2l},
\label{eq:leg_pol}
\end{equation}
where $\nu=l_1+l_2-m_1-m_2$ and $\big [\cdot\big]$ means the integer
part.  In this way, for certain $l$ we may seek for all the products
$P^{m_1}_{l_1}P^{m_2}_{l_2}$ for which $P^{m_1+m_2}_{l_1+l_2-2l}$
appears in the linear combination of Eq. (\ref{eq:leg_pol}). We are
thus searching for modes $(m_i,l_i)$, $i=1,2$, which ``feed'' -by
means of the products $P^{m_1}_{l_1}P^{m_2}_{l_2}$- the mode
$(m,l)=(m_1+m_2,l_1+l_2-2l)$. Some of the found modes $(m_i,l_i)$ are
in resonance with the ``feeded'' mode $(m,l)=(m_1+m_2,l_1+l_2-2l)$.

The linear combinations of Eq.(\ref{eq:leg_pol}) are tabulated in the
appendix of \cite{Hwa95}, up to degree $5$, and help us to exemplify
the relation between resonances and nonlinear interaction of the
modes. For example, by looking for the products which feed
$(m,l)=(3,3)$ we find
\begin{eqnarray}
  P^2_2P^1_1=g(P^3_3),~P^2_3P^1_2=g(P^3_5,P^3_3),
  \label{eq:rel1}
\end{eqnarray}
where $g$ is a linear function different for each product. One of
these relations provide the resonance condition
$f^3_3=f^1_1+f^2_2$. We note that we have excluded products such as
\begin{eqnarray*}
P^1_3P^2_2=g(P^3_5,P^3_3),~P^1_4P^2_3=g(P^3_7,P^3_5,P^3_3),
\end{eqnarray*}
since they are equivalent to those of Eq.(\ref{eq:rel1}) in the sense
that for the modes $(1,1)$ and $(1,3)$ (resp. $(1,2)$ and $(1,4)$) the
same frequency can be selected $f^1_1=f^1_3$ (resp. $f^1_2=f^1_4$), as
they both represent the equatorially symmetric (resp. equatorially
antisymmetric) part of the $m=1$ mode. We have also not considered the
coupling with the $m=0$ mode.

\begin{figure}
  \hspace{0.mm}\includegraphics[width=0.95\linewidth]{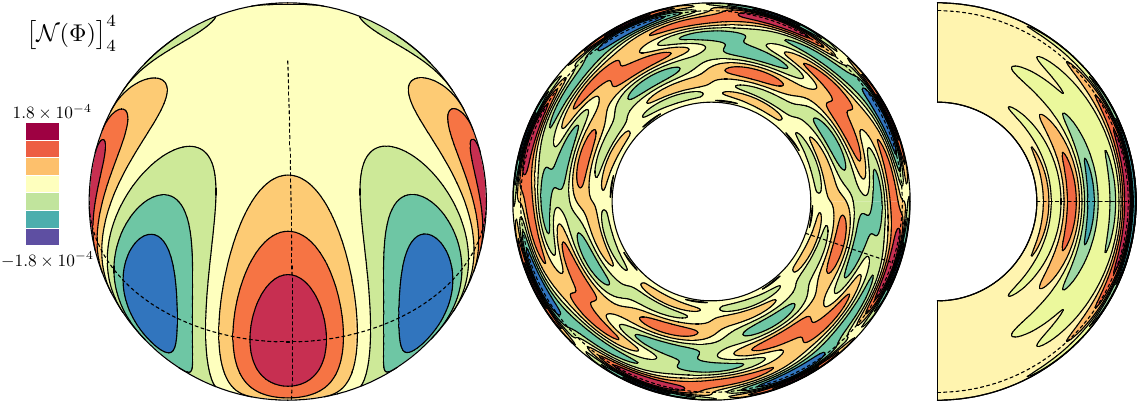}\\
  \hspace{0.mm}\includegraphics[width=0.95\linewidth]{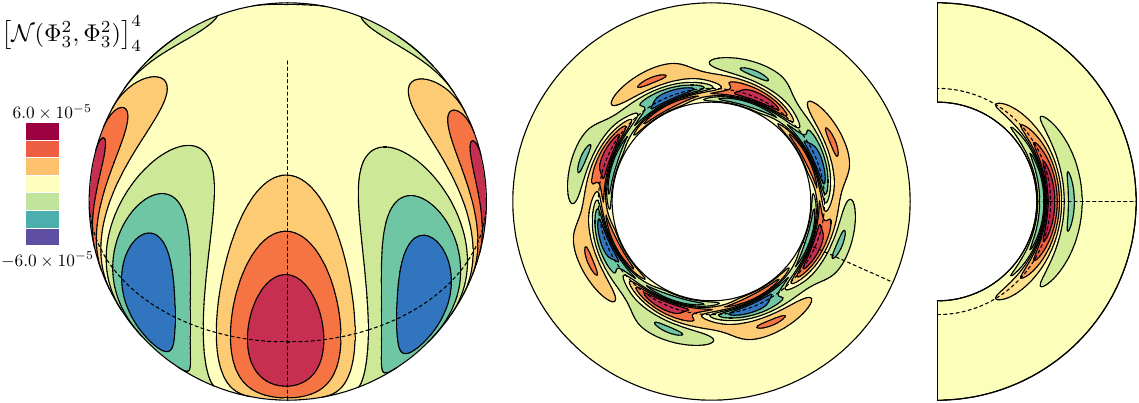}\\
  \hspace{0.mm}\includegraphics[width=0.95\linewidth]{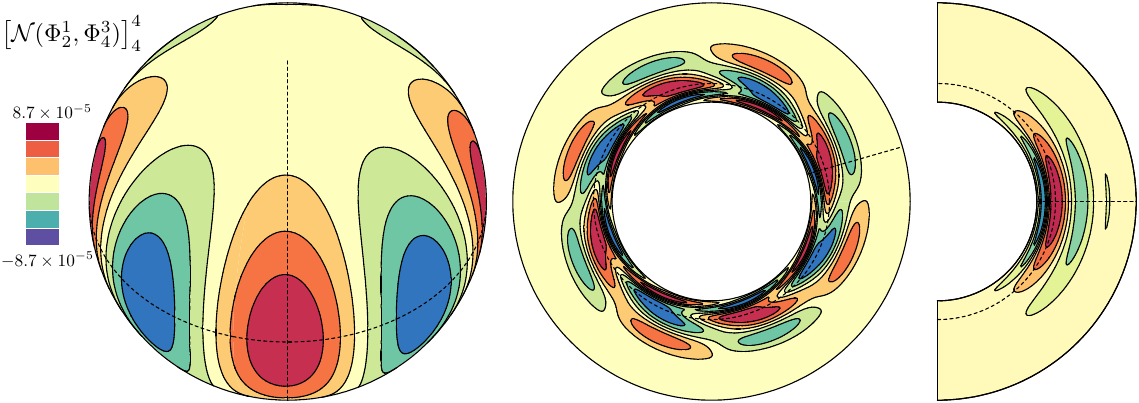}\\
\caption{As figure \ref{fig:ctplm3_m3l3} but for the $(m,l)=(4,4)$
  component of the nonlinear term. In this figure the modes
  $(m,l)=(2,3),(2,3)$ and $(m,l)=(1,2),(3,4)$ are considered in the
  middle and bottom row.  The solution is a 2T MRW with $m=1$
  azimuthal symmetry at $\Ha=3.5$.}
\label{fig:ctplm3_m4l4}
\end{figure}

\begin{figure}
  \hspace{0.mm}\includegraphics[width=0.95\linewidth]{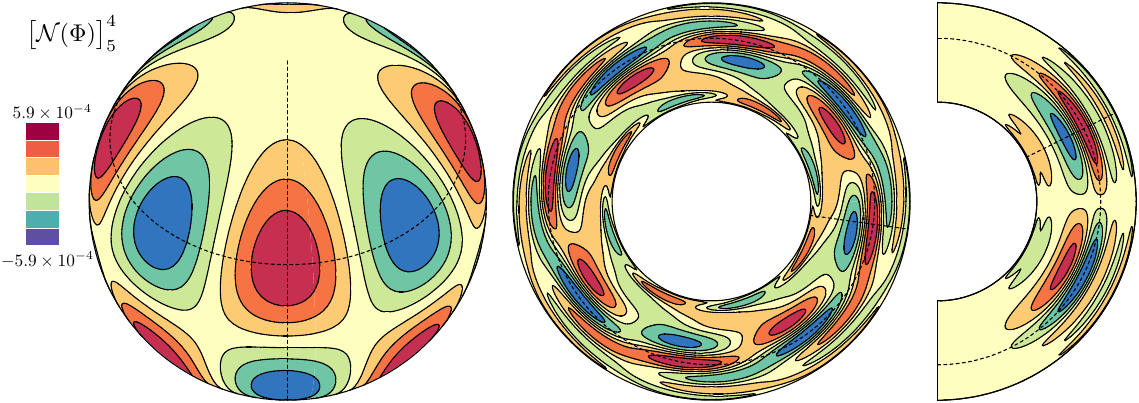}\\  
  \hspace{0.mm}\includegraphics[width=0.95\linewidth]{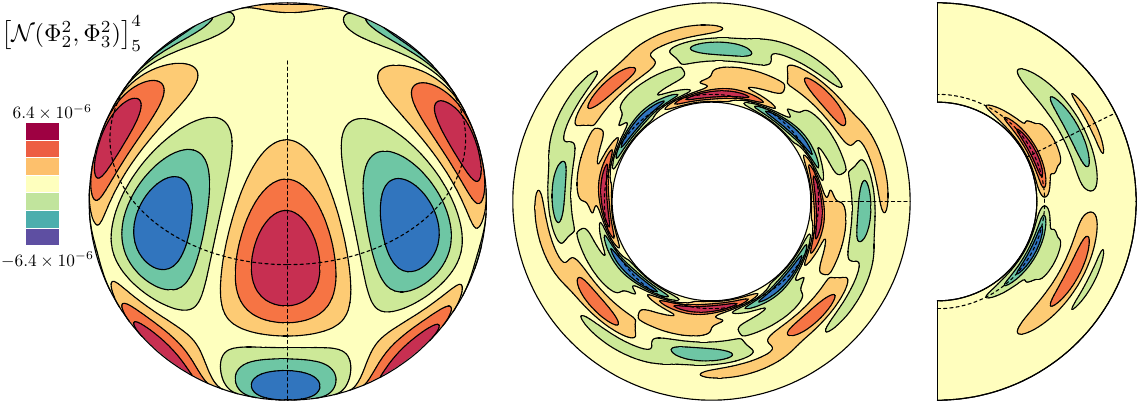}\\
  \hspace{0.mm}\includegraphics[width=0.95\linewidth]{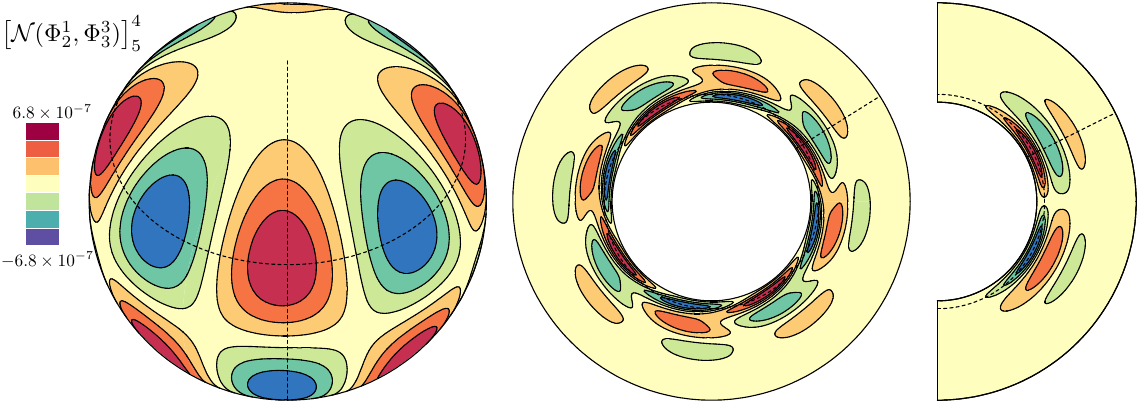}\\
  \hspace{0.mm}\includegraphics[width=0.95\linewidth]{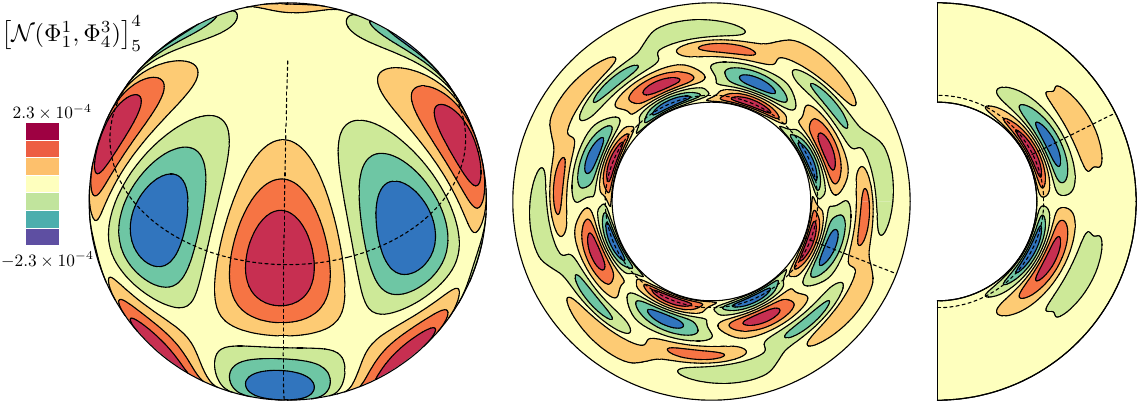}\\  
\caption{As figure \ref{fig:ctplm3_m3l3} but for the $(m,l)=(4,5)$
  component of the nonlinear term. In this figure the modes
  $(m,l)=(2,2),(2,3)$, the modes $(m,l)=(1,2),(3,3)$, and the modes
  $(m,l)=(1,1),(3,4)$, are considered in the 2nd, 3rd, and 4th rows,
  respectively. The solution is a 2T MRW with $m=1$ azimuthal
  symmetry at $\Ha=3.5$.}
\label{fig:ctplm3_m4l5}
\end{figure}

In case of the $(m,l)=(3,4)$ mode the situation is analogous and the
products feeding this mode are
\begin{eqnarray*}
  P^2_2P^1_2=g(P^3_4),~P^2_3P^1_1=g(P^3_4),~P^2_3P^1_1=g(P^3_4),
\end{eqnarray*}
for which both resonance relations have been found (see Table
\ref{table:res_2T_m1}) in the scenario studied in
\S\ \ref{sec:s_wav}. For the azimuthal wave number $m=4$ the products
feeding the $(4,4)$ mode are
\begin{eqnarray*}
  P^2_2P^2_2=g(P^4_4),~P^2_3P^2_3=g(P^4_6,P^4_4),\\
  P^3_3P^1_1=g(P^4_4),~P^3_4P^1_2=g(P^4_6,P^4_4)
\end{eqnarray*}
and for the $(4,5)$ mode are
\begin{eqnarray*}
P^2_3P^2_2=g(P^4_5),~P^3_3P^1_2=g(P^4_5),~P^3_4P^1_1=f(P^4_5).
\end{eqnarray*}
As in the case for the $m=3$ wave number the resonances found for the
$(4,4)$ and $(4,5)$ modes arise from the above relationships. We have
also checked the case $m=5$.

To further investigate the coupling between the resonant modes for the
poloidal field we compute its associated nonlinear term from the
Navier-Stokes equation, i.\,e.,
\begin{equation}
\mathcal{N}(\ve)=\Ree~\rv\cdot\nabla\times\nabla\times {\bf
  \vo\times\ve},
\label{eq:nl_pol}
\end{equation}
but with only considering the poloidal component when computing the
velocity $\ve$, and vorticity $\vo$ fields (i.\,e. we set the toroidal
field to zero $\Psi\equiv 0$), so $\mathcal{N}\equiv
\mathcal{N}(\Phi)$.  We then compare the spatial patterns of the
scalar $\mathcal{N}(\Phi)$ obtained by using all the modes in the
spherical harmonic expansion of the poloidal field
(Eq.~(\ref{eq:serie_phi})), with the patterns of $\mathcal{N}(\Phi)$
obtained from a poloidal field with only nonzero amplitudes for a
selected set of two resonant modes. As it is shown in the following,
the set of two resonant modes provides a first order approximation of
the patterns obtained with the full spherical harmonic expansion. For
the pattern comparison we have considered the 2T MRW with $m=1$
azimuthal symmetry at $\Ha=3.5$, the resonances of which have been
studied in \S\ \ref{sec:s_wav}, concretely in table
\ref{table:res_2T_m1}.

The first row of figure \ref{fig:ctplm3_m3l3} displays the contour
plots of $\big[\mathcal{N}(\Phi)\big]^3_3$, i.\,e., the spherical
harmonic amplitude $(m,l)=(3,3)$ of the nonlinear term given in
Eq. \ref{eq:nl_pol}, on spherical, meridional and latitudinal sections
(from left two right). The position of the sections, marked with
dashed lines, matches the position of a local maxima. The second row
shows the corresponding contour plots of
$\big[\mathcal{N}(\Phi)\big]^3_3$ for the case that only the
amplitudes $(m,l)=(1,1),(2,2)$ of the poloidal field are non-zero. The
similarity between the radial structure among the two rows is a clear
indication that the nonlinear interaction of the $(1,1)$ and
$(2,2)$ resonant modes is sufficient to provide a first order
approximation of the nonlinear term feeding the equatorially symmetric
$(3,3)$ mode. Figure \ref{fig:ctplm3_m3l4} summarises the comparison
in the case of the modes which are resonant with the equatorially
antisymmetric mode $(3,4)$ (shown in the first row). These are the
modes $(1,2),(2,2)$ and $(1,1),(2,3)$ displayed in the middle and
bottom row, respectively.

The patterns of the nonlinear terms in the case of the resonant modes
with azimuthal wave number $m=4$ are displayed in
Fig. \ref{fig:ctplm3_m4l4}, for the equatorially symmetric $(4,4)$
mode, and in Fig \ref{fig:ctplm3_m4l5}, for the equatorially
antisymmetric $(4,5)$ mode, respectively. The patterns for the
resonant modes $(2,3),(2,3)$ and $(1,2),(3,4)$ and those obtained from
the full spherical harmonics expansion match only close to the inner
boundary, so non-resonant modes should be considered to approximate
the nonlinear term close to the outer boundaries. For the equatorially
antisymmetric mode $(4,5)$ the resonant modes provide a reasonable
approximation of the nonlinear term.

\section{Conclusions}
\label{sec:conc}

A comprehensive frequency analysis of several equatorially symmetric
and antisymmetric poloidal modes has been presented for different
types of weakly magnetic spherical Couette flows, in the context of
the radial jet instability (\cite{HJE06,TEO11,GaSt18}). Two different
bifurcation scenarios have been considered to evidence the mechanisms
for the development of resonant triads among modes $(m,l)$, having
different spherical harmonic order $m$ and degree $l$.

The first scenario comprises the first, second and third bifurcations
from the base state giving rise to stable flows. The second scenario
involves up to five bifurcations giving rise to either unstable or
stable flows. In both scenarios the sequence of bifurcations includes
the path:
$$\text{base
  state}\rightarrow\text{RWs}\rightarrow\text{2T-MRWs}\rightarrow\text{3T-MRWs}$$
for which the right arrows indicate Hopf bifurcations, which may break
the azimuthal symmetry of the flow. The above sequence of bifurcations
is generic in {\bf{SO}}$(2)\times${\bf{Z}}$_2$-equivariant systems
when the first bifurcation breaks the axisymmetry of the base
state. In this case, RWs are the first time dependent solutions
(e.\,g. \cite{Ran82,CrKn91,GLM00}).

The type of time dependence of a RW, for which the main time scale is
associated to a solid body rotation of a flow pattern, allows to
define $\overline{\varphi}=\varphi+\omega t$, with $\omega$ being the
rotation frequency of the wave, so the mathematical description in
terms of spherical harmonics become equivalent to solutions of the
inviscid equation (\cite{Gre68}). This is also true for MRW which are
quasiperiodic flows with two incommensurate frequencies for which, due
to the nonlinear coupling, resonance relations can be defined
following similar arguments as those given in \citet{BTHW18}.

For all the MRW explored, which have the azimuthal symmetry broken, a
multitude of resonance relations, both in terms of spherical harmonic
order $m$ and degree $l$, are found. We have considered modes of the
type $(m,m)$ and $(m,m+1)$ to investigate the role of the equatorially
symmetric and antisymmetric part of the flow. In addition, we have
analyzed the nonlinear interactions giving rise to triadic resonances
in terms of Legendre polynomials.

The phase coupling among resonant modes, pointed out in \cite{HHT16}
and further investigated in \citet{BTHW18}, is a consequence of the
regularity of MRW, as they are quasiperiodic solutions. For this type
of solutions, frequency spectra do not depend on time (\cite{GSGS21}).
In addition, triadic resonances are observed for chaotic flows
originated from further bifurcations of MRWs. This is because the main
two frequencies of these chaotic flows are very similar to those of
the MRW, from which they bifurcate (\cite{GSGS20,GSGS20b}). Because of
this, triadic resonances are very common for the radial jet
instability of the MSC flow.

The results of the present study are compared with previous studies
devoted to triadic resonances in spherical systems. In the case of
spherical Couette flow resonant modes have been studied in
\citet{BTHW18} and for rotating thermal convection in full spheres
resonant solutions are investigated in \cite{Lin21}. In contrast to
these studies, for which all the free resonant modes are inertial and
the base mode is strongly geostrophic, our simulations demonstrate the
existence of resonances between modes which are neither geostrophic
nor inertial, but related with the radial jet instability.

The existence of triadic resonances in spherical systems can thus be
explained in terms of bifurcation theory since they are associated to
the existence of MRW. In this framework, we argue that an inertial
regime is not a requirement for the development of resonant triads as
our results point out. We remark that in spherical systems the
interpretation of resonances in terms of MRW, i.\,e. equivariant
dynamical systems theory, is not in contradiction with the
interpretation in terms of inertial waves since MRWs in the inertial
regime have been found in the case of rotating thermal convection
(e.\,g. \cite{GCW19,Lin21}).

We finally note that MRW are common in rotating convective spherical
shells (e.\,g. \cite{GSDN15,GNS16}) in a regime in which the onset of
convection is of columnar type (\cite{Zha92}). We have checked that
for these type of MRW resonant triads also occur.

\begin{acknowledgments}
This project has received funding from the European Research Council
(ERC) under the European Union’s Horizon 2020 research and innovation
programme (grant agreement No 787544). 
\end{acknowledgments}


\section*{Data availability}
The data that support the findings of this study are available from
the corresponding author upon reasonable request.


\renewcommand\refname{References}

%

\end{document}